\documentclass[10pt,twocolumn,aps,prx,floatfix,superscriptaddress,nofootinbib]{revtex4-2}
\usepackage{amsmath}
\usepackage{amsfonts}
\usepackage{amssymb}
\usepackage{graphicx}
\usepackage{color}
\usepackage{bm}
\usepackage{dsfont}
\usepackage{soul}
\usepackage{mathtools}
\usepackage[normalem]{ulem}
\usepackage{multirow}
\usepackage{wasysym}

\DeclarePairedDelimiterX\braket[2]{\langle}{\rangle}{#1 \delimsize\vert #2}
\usepackage[final]{hyperref} 
\hypersetup{
	colorlinks=true,       
	linkcolor=blue,        
	citecolor=blue,        
	filecolor=magenta,     
	urlcolor=blue         
}
\newcommand{\beq}{\begin{equation}}
\newcommand{\eneq}{\end{equation}} 
\newcommand{\be}{\begin{equation}}
\newcommand{\ee}{\end{equation}}
\newcommand{\bea}{\begin{eqnarray}}
\newcommand{\eea}{\end{eqnarray}}

\begin{document}

\title{Probing nonequilibrium transport and density profiles in tilted interacting fermion chains with local reservoirs}

\author{Andrea Nava}
\affiliation{Institut f\"ur Theoretische Physik, Heinrich-Heine-Universit\"at, 40225 D\"usseldorf, Germany}

\author{N. Aucar Boidi}
\affiliation{The Abdus Salam International Centre for Theoretical Physics, Strada Costiera 11, I-34151, Trieste, Italy}

\author{Reinhold Egger}
\affiliation{Institut f\"ur Theoretische Physik, Heinrich-Heine-Universit\"at, 40225 D\"usseldorf, Germany}

\author{Karen Hallberg}
\affiliation{Centro Atómico Bariloche, CNEA, Bariloche, Argentina}
\affiliation{Instituto Balseiro and Instituto de Nanociencia y Nanotecnología, CNEA-CONICET, Bariloche, Argentina}

\author{C. R. Proetto}
\affiliation{Centro Atómico Bariloche, CNEA, Bariloche, Argentina}
\affiliation{Instituto Balseiro and Instituto de Nanociencia y Nanotecnología, CNEA-CONICET, Bariloche, Argentina}

\begin{abstract}
We investigate how local particle reservoirs can probe transport and density profiles in tilted spinless fermion chains with nearest-neighbor interactions. Using finite-chain Lindblad calculations with particle injection and extraction at opposite ends, we define a threshold-dependent transmission range from the stationary current as the chain length is varied. This range decreases with increasing tilt and follows the trends of the equilibrium density-crossover width in the parameter regimes studied. Interactions can enhance transmission by compensating field-induced energy costs, whereas strongly polarized configurations suppress transport. A third, weakly coupled extraction reservoir provides a complementary spatial probe: its current is exactly proportional to the local occupation in the presence of the probe, and the normalized current approaches the unperturbed nonequilibrium density as the probe coupling vanishes. We illustrate the detection of staggered density modulations and show that bulk densities can continue to relax after boundary currents have nearly stabilized. These protocols connect current measurements to spatial transmission and density structure in driven interacting chains.
\end{abstract}

\maketitle

\section{Introduction}
\label{intro}

Spatial localization provides a mechanism for suppressing transport in lattice systems even in the absence of an
ordinary band gap. In noninteracting disordered systems, interference between scattering paths can prevent the
spreading of a wave packet, giving rise to Anderson localization \cite{anderson58}. A conceptually different mechanism occurs in a clean lattice subject to a static linear electric potential. The applied electric field induces Bloch oscillations and prevents transport, leading to what is now known as Wannier-Stark localization. The corresponding Wannier-Stark eigenstates are spatially localized. For an infinite nearest-neighbor
chain, their energies form an equally spaced ladder and their amplitudes are expressed through Bessel functions,
with a characteristic spatial extent set by the ratio of hopping to field strength \cite{WSloc1,WSloc2,Boidi_2025}.
Wannier-Stark physics has been observed in semiconductor heterostructures \cite{mendez88} and was explored in detail in controllable platforms such as cold atomic gases \cite{wilkinson1996,geiger2018} and superconducting quantum
processors \cite{guonpj}. Such platforms allow the potential gradient and the coherent dynamics to be varied independently, making tilted lattices useful settings for investigating the relation between spatial confinement and transport.

Interactions considerably enrich this picture. In disordered systems, their interplay with localization
can inhibit thermalization and preserve information about the initial state, triggering the intensely studied field of
many-body localization physics starting with Refs.~\cite{gornyi2005,basko2006}.
Strong potential gradients can also produce slow relaxation and persistent density inhomogeneities in \emph{clean interacting systems}, the case studied also in this work.   
Early contributions identified signatures of a disorder-free localized regime, commonly termed Stark
many-body localization \cite{Schulz2019,nieuw2019}.
Subsequent work investigated how these signatures depend
on the potential profile, including the addition of
harmonic confinement \cite{Yao2021}.
Experiments with trapped ions observed persistent memory
in a spatially increasing field \cite{Morong2021}, while
tilted Fermi-Hubbard chains realized with ultracold atoms
revealed nonergodic dynamics associated with kinetic
constraints \cite{Scherg2021} and explored regimes of
Hilbert-space fragmentation \cite{Kohlert2023}.

The interpretation of slow dynamics in tilted interacting
systems nevertheless requires care. A strong potential gradient generally suppresses particle transport by making individual hops
energetically costly. In interacting chains,
however, collective rearrangements can permit
motion on much longer timescales. Indeed,
transport studies have found that the time
before appreciable spreading occurs can grow
rapidly with field strength and system size
\cite{Zisling2022}. A chain can therefore appear
localized over a long observation window without
transport being strictly absent at later times.
Other work relates slow density
relaxation to emergent dipole-moment conservation and
subdiffusion \cite{Nandy2024}. The precise interaction type
also matters. For nearest-neighbor density interactions,
the terms that redistribute particles between
Wannier-Stark orbitals become weak at strong tilt,
although interaction-induced energy shifts remain
important \cite{Krajewski_2025}.
These results motivate a distinction between spatial
confinement on accessible scales, long relaxation times,
and the existence of an asymptotic many-body localized
phase. They also motivate a characterization of the
distance over which a measurable current can
be sustained under a specified driving protocol. 

A complementary view of confinement follows from
equilibrium density profiles. In a tilted chain, the
ground-state occupation changes from an almost filled
region at low potential to an almost empty region at
high potential. For spinless fermions with nearest-neighbor
repulsion, interactions broaden the intermediate-density
region and can produce a central charge-density-wave
(CDW) pattern \cite{Boidi_2025}. The width of this
crossover reflects competition between the potential
gradient, hopping, and interaction energies.
It provides a useful spatial characterization of the
closed system, but its relation to transport through an
open, continuously driven chain is not immediately transparent.
The reservoirs select occupations through their coupling
to the system, and a stationary current depends on both
the available states and the processes connecting them.

Coupling a tilted system to reservoirs changes the
physical question from the spreading of an isolated
excitation to sustained particle transmission.
The distinction is already relevant without interactions:
Bloch oscillations in an isolated lattice can coexist
with transient oscillations that decay toward a finite
stationary current when a finite device is connected
to leads \cite{Pinho2023}.
Reservoir-based approaches have also been used to study
transport through periodically driven tilted lattices,
where the reservoir relaxation must be controlled to
recover the response of continuum leads \cite{De2023}.
The transport properties of a contacted system therefore
depend on how particles enter and leave, as well as on
the confinement generated by its Hamiltonian.

The celebrated Lindblad master equation provides a widely used
framework for addressing these questions when reservoir
memory effects can be neglected and system-reservoir couplings are weak (Born-Markov approximation) \cite{Lindblad_1976,Breuer2007}.
The Lindblad master equation combines coherent Hamiltonian evolution with
dissipative processes specified by jump operators.
For boundary-driven spin and fermion chains, it gives
access to the approach to a nonequilibrium steady state
(NESS), its density profile, and its stationary currents
\cite{Landi_2022}. Matrix-product simulations established
practical methods for computing such states
\cite{Prosen2009}, while analytical constructions for
the boundary-driven XXZ chain provided exact stationary
solutions and constraints on transport
\cite{Prosen_2011,Prosen_2011_2,Karevski_2013}.
These developments complement the broader theory of
transport in one-dimensional (1D) quantum systems
\cite{Jin_2020,Bertini_2021}. Under suitable conditions, the
connection between boundary-driven dynamics and
equilibrium correlation functions also allows
stationary transport to be related to properties
of the corresponding closed system
\cite{Heitmann_2023}.

Far from equilibrium, the reservoirs can select
configurations that differ substantially from those
relevant to linear response or low-temperature
equilibrium states. Strong driving in interacting fermion chains
can generate polarized domains and negative
differential conductivity
\cite{Benenti_2009,Nava_2021} and modify transport in a symmetry-dependent manner \cite{Visuri_2023,Gievers_2024}.
Dissipation need not simply degrade transport.
Dephasing can release interaction-induced blocking
and enhance the current \cite{Mendoza_2013,Lee2022,Jacob_2024}.
Related Lindblad approaches have used stationary
transport and reservoir-induced dynamics to probe
localization and topology in disordered Su-Schrieffer-Heeger (SSH) and
Kitaev chains
\cite{Carmele2015,Dabbruzzo2021,Nava_2023,Yan2023,Cinnirella_2024,Scocco2024,Cinnirella_2025,Berger_2025,
Cinnirella_2026}.  One can also study the flow of quantum information in 1D fermion 
chains using this framework, see Ref.~\cite{Nava2026a}.
Together, these studies establish local reservoirs
as active elements in both preparing and probing
quantum states.  

For interacting tilted fermion chains, the competition between
driving, confinement, and interactions is particularly
pronounced. The electric field and the strong interaction can each
suppress particle transmission, yet their combination can enhance
it if a change in interaction energy compensates the
potential cost of particle motion.
This competition
produces strong rectification in boundary-driven tilted
systems. Driving particles uphill can reveal conduction
resonances associated with avoided crossings of many-body
levels and enhanced coherences, whereas the opposite
driving direction can remain strongly insulating
\cite{Mendoza_2024}.
Thus a suppressed current does not identify its mechanism
uniquely, and increasing the interaction need not
monotonically strengthen localization.
The spatial density profile and the transmitted current
provide complementary information about the configurations
selected by the drive.

The reservoir model must also be specified when comparing
these observables. The relation between local dissipators,
energy-resolved transitions, and thermalization is
nontrivial in interacting systems. Developments beyond
the full secular approximation address thermodynamic
consistency \cite{Potts2021}, while derivations based on
the Lieb-Robinson bound clarify conditions for local
many-body master equations \cite{Shiraishi_2025}.
Here we adopt an energy-independent local gain-loss
description, appropriate to effectively filled and
empty reservoirs over the relevant transition window,
or to engineered Markovian injection and extraction.
The reservoirs allow for particle exchange at selected (``local'')
sites without requiring the chain to approach a Gibbs
state. Thermal equilibrium is introduced separately
as a benchmark for the spatial structures favored
by the Hamiltonian.

An experimentally important question is how spatial information can be obtained from such an open quantum system.
A current measured at a single fixed contact separation reveals whether particles are transmitted between a pair of
reservoirs, but it does not by itself determine the distance over which transmission
is appreciable. Conversely, reconstructing every local
occupation number provides detailed spatial information but
requires position-resolved measurements. Spatially controllable reservoirs could connect those two
types of observation. Changing the length of the contacted segment probes the range of transmission,
while a weakly coupled local probe reservoir can convert the local occupation 
into a measurable particle current between the system and the auxiliary reservoir.
The central objective of this work is to develop these complementary uses of local reservoirs for
tilted interacting fermionic chains.  A mobile local reservoir could, for instance, be engineered by means
of a scanning tunneling microscope (STM) tip operated in a limit where it represents an effectively empty broad-band extraction reservoir over the relevant removal spectrum, with a calibrated local coupling.

First, we define a penetration length from the largest chain length whose stationary current exceeds a predesignated
detection threshold. We here implement this measurement
numerically by varying the total length of a chain
with particle-injecting and -extracting end-point reservoirs, respectively.
In this process, we 
keep the field, the interaction strength, and the contact rates fixed.
This construction directly measures transport through
an isolated segment and avoids simulating additional
exterior sites for the interacting problem.
The resulting length characterizes the transmission properties for
the chosen contacts and the current resolution.
Remarkably, our finite-chain results show a close correspondence
with the equilibrium density-crossover width, suggesting
that the same competition between field and interaction
energies organizes both spatial scales over the
parameter regimes studied below. 
The current scans also reveal interaction-assisted transmission and its eventual
suppression when strongly polarized configurations develop.

Second, we retain the end-point drive and attach another weakly coupled reservoir at a tunable bulk
site. Its outgoing current equals the extraction
rate times the local occupation in the state with
the probe attached. This identity is exact for both
noninteracting and interacting chains.
In the weak-probe limit, the normalized current
approaches the occupation of the unperturbed NESS,
providing a current-based readout of its spatial
density profile. We use this relation to investigate
the detection of staggered density modulations
associated with CDW precursors in finite chains.
The time-dependent occupations further show that
bulk redistribution can remain slow after boundary
currents have nearly stabilized, making the preparation
time relevant to the interpretation of a spatial scan \cite{Wolff_2020}.

Throughout, we compare the driven profiles with a
thermal equilibrium steady state (TESS).
The two density distributions can become remarkably
similar even though their state-selection mechanisms
differ. The thermal benchmark favors low-energy
configurations, whereas the local reservoirs populate
and deplete states through their contact matrix
elements. Symmetry relations help organize this
comparison and explain how reversing the field and
interaction can connect NESS profiles resembling
low-energy equilibrium configurations to profiles
characteristic of the upper spectral edge.
These correspondences concern the measured densities
and do not require the driven state to be thermal.
The combined analysis therefore identifies what can
be learned from studies of transport over different length scales
and which types of additional spatial information become
accessible through the use of a mobile local reservoir.

The remainder of this paper is organized as follows.
Section~\ref{sec2} introduces the tilted
interacting fermion chain model, the local reservoirs, the thermal
benchmark, and the numerical method.
In Sec.~\ref{sec3}, we compare stationary density
profiles obtained from the NESS solution of the Lindblad equation to 
the TESS reference state and provide physical interpretations of our 
results.
In Sec.~\ref{sec4}, we then
examine the relaxation dynamics and
interaction-dependent transport. In particular, we extract a
penetration length by varying the length of chains
coupled to end-point reservoirs.
We also investigate the formation of
staggered density profiles and their detection
through a weakly coupled mobile extraction tip.
We present our conclusions in Sec.~\ref{sec5}.
Appendix~\ref{appA} develops the
energy-eigenbasis description of reservoir coupling
and interaction-assisted transport, while
Appendix~\ref{symmetries} derives symmetry
relations governing the spectra and the thermal
and driven density profiles.  Throughout the paper, we use
units with $\hbar=k_B=1$.

\section{Model and dissipative probes}
\label{sec2}

In this section, we introduce the model and the Lindblad master equation approach employed to solve for the dynamics of the system state $\rho(t)$.  In Sec.~\ref{sec2a}, we introduce the Hamiltonian and the Lindblad equation.
The jump operators used in this work are explained in Sec.~\ref{sec2b}, along with a discussion of the density profile and the currents flowing in the chain.  In Sec.~\ref{sec2c}, we describe a thermal equilibrium reference state to which the 
full nonequilibrium solution will be compared later on.  Finally, in Sec.~\ref{sec2d} we present details on the numerical solution of the Lindblad equation.

\subsection{Tilted interacting fermion chain}\label{sec2a}

\begin{figure}
\center
\includegraphics*[width=\linewidth]{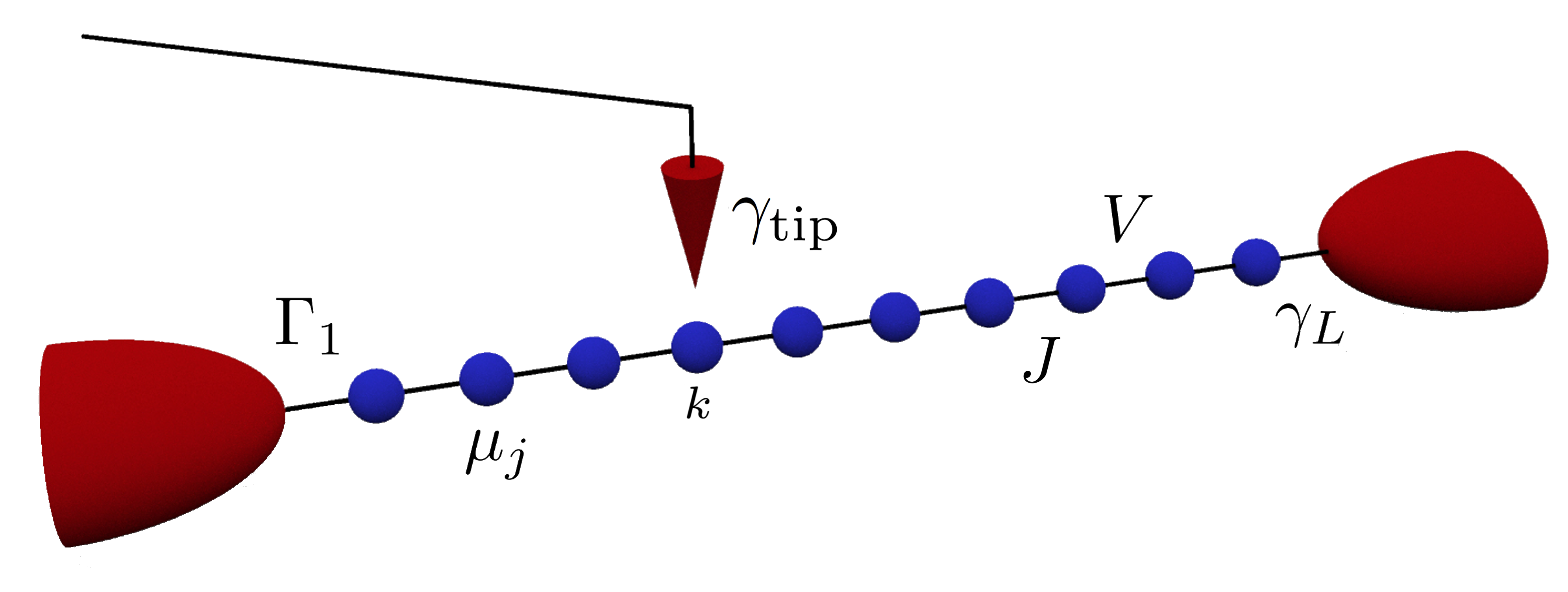}
\caption{Schematic sketch of the 1D tilted fermion chain with $L$ sites described by the Hamiltonian \eqref{eq:Ham} and coupled by jump operators to end-point reservoirs (shown in red), with rate $\Gamma_1$ at site $j=1$ and rate $\gamma_L$ at site $j=L$. The left reservoir injects particles at site $j=1$, while the right reservoir extracts them at site $j=L$, thereby establishing a NESS. In addition, a mobile tip (red cone)
is weakly coupled to site $k$ with rate $\gamma_{\mathrm{tip}}$. 
 This additional extraction reservoir allows one to scan the local density profile by measuring the respective particle current flowing from the system to the tip.}
\label{fig:fig1}
\end{figure}

We consider a 1D fermion chain with $L$ sites and open boundary conditions. The geometry is schematically sketched in Fig.~\ref{fig:fig1}. Denoting the nearest-neighbor hopping amplitude by $J$, using a linear potential tilt $\mu_j$, and nearest-neighbor density-density interactions of strength $V$, the system Hamiltonian 
is given by 
\begin{align}
H = & -J \sum_{j=1}^{L-1} \left(c_{j+1}^{\dagger}c_j^{}+c_j^{\dagger}c_{j+1}^{}\right)
+\sum_{j=1}^{L}\mu_j n_j \nonumber\\
&+V\sum_{j=1}^{L-1}\left(n_j-\frac12\right)\left(n_{j+1}-\frac12\right).
\label{eq:Ham}
\end{align}
Here, $c_j$ is a fermion annihilation operator at lattice site $j$, with $n_j=c_j^{\dagger} c_j^{}$ denoting the corresponding fermion number operator. We assume a linear potential tilt,
\begin{equation}
\mu_j=eaF(j-j_c),
\label{eq:potential}
\end{equation}
where $e$ is the electron charge, $a$ the lattice spacing, $F$ the potential slope which is proportional to the strength of the electric field, and $j_c=\frac{L+1}{2}$ the point at which the potential vanishes, $\mu_{j_c}=0$. 
The $\frac12$ shifts in the interaction term energetically favor the half-filled configuration.
From now on, we use units with $e=a=1$ such that $F$ denotes the energy drop per bond, and measure 
energies and inverse times in units of $J$. 

To study the dynamics of the open quantum system obtained by including contacts to reservoirs, one can consider the effective dynamics of the system by integrating  over the reservoir degrees of freedom. In doing so, we resort to the so-called Born-Markov approximation,   neglecting memory effects under the assumption that the bath relaxation time is much shorter than the characteristic timescales of the system. In addition, one assumes that all
system-reservoir couplings are weak. The reduced density matrix $\rho$ of the system then obeys the Lindblad master equation \cite{Breuer2007}
\begin{equation}
\dot{\rho}(t)=\mathcal L[\rho(t)] =-i[H,\rho(t)]+\sum_k\mathcal D[L_k]\rho(t),
\label{eq:lindbladeq}
\end{equation}
with
\begin{equation}\label{superop}
\mathcal D[L]\rho=L\rho L^\dagger-\frac12\{L^\dagger L,\rho\}.
\end{equation}
Here, the first term in the Liouvillian ${\cal L}$ describes unitary evolution as determined by $H$. The second term, the so-called Lindbladian, includes dissipation and decoherence due to the system-reservoir couplings.
This contribution depends on a set of jump operators $\{L_k\}$.

\subsection{Local reservoirs and mobile probe protocol}\label{sec2b}

A reservoir coupled only to site $j$ injects (``gain")  or extracts (``loss") fermions at this site with given
rates. We model these processes through Lindblad jump operators which are
linear in the respective local fermion operators,
\begin{equation}
L_{\mathrm{in},j}=\sqrt{\Gamma_j} \, c_j^\dagger,
\quad
L_{\mathrm{out},j}=\sqrt{\gamma_j}\, c_j^{},
\label{eq:in_out}
\end{equation}
with the rates $\Gamma_j$ and $\gamma_j$ for particle injection and extraction from the system, respectively.
We model each local reservoir as particle source and/or sink, where the injection and extraction rates are assumed not to depend on the system transition energies. For a thermal fermionic reservoir, this approximation requires both the reservoir spectral density and the Fermi distribution to be approximately constant over the relevant energy range. These conditions can be realized in the large-bias limit, where in effect the injecting reservoir is filled and the extracting reservoir is  empty throughout the relevant transition window \cite{Benenti_2009,Nava2026a}. Alternatively, energy-independent rates can describe engineered Markovian injection and loss
processes. We parametrize the rates as
\begin{equation}
\Gamma_{j}=g_j f_j, \quad\gamma_{j}=g_j(1-f_j),
\label{eq:rates}
\end{equation}
where $g_j=\Gamma_{j}+\gamma_{j}$ sets the overall coupling rate and $0\leq f_j\leq1$ specifies
the occupation factor imposed by reservoir $j$. In the large-bias configuration considered below,
$f_{j=1}=1$ and $f_{j=L}=0$, such that the left reservoir only injects particles and the right reservoir only extracts them. In the three-reservoir configuration, these two contacts remain fixed and an additional tip is attached at a variable bulk site $k$.  At sites
not coupled to any reservoir, we have $\Gamma_j=\gamma_j=0$.
We emphasize that ``large-bias conditions'' refers to the external reservoirs and not the electric field parameter $F$. Throughout the paper, we refer to $F$ as field parameter.  With suitable gate electrodes, one can apply an external electric field independent of the bias voltage applied through the reservoirs.   

Once the density matrix $\rho(t)$ has been obtained by solving Eq.~\eqref{eq:lindbladeq}, the time-dependent expectation value of any observable $O$ follows as
$\langle O\rangle_t=\mathrm{Tr}[O\rho(t)]$.
The local density is described by the expectation value
\begin{equation}
\bar n_j(t)=\langle n_j\rangle_t=\mathrm{Tr}\!\left[n_j\rho(t)\right].
\label{eq:local_density}
\end{equation}
To define the particle current, we use the continuity equation, i.e., the equation of
motion for $\bar n_j(t)$. The bond-current operator is
\begin{equation}
J_{j,j+1} = -iJ\left(c_{j}^\dagger c^{}_{j+1}-c_{j+1}^\dagger c^{}_{j}\right),
\label{eq:bond_current}
\end{equation}
where positive current denotes transport from site $j$ to $j+1$. The potential and density-density interaction terms commute with $n_j$ and therefore do not contribute directly to the continuity equation.
The reservoir contributions follow from the adjoint Lindblad dissipator $\mathcal D^\dagger[L]O$. 
Injection at site $j$ contributes the term $\Gamma_j(1-\bar n_j)$, whereas extraction contributes
$-\gamma_j\bar n_j$. It is convenient to introduce a local source operator,
\begin{equation}
J_{E,j} =\Gamma_j(1-n_j)-\gamma_j n_j = g_j(f_j-n_j),
\label{eq:environment_current}
\end{equation}
whose expectation value is the net particle current entering the chain from the reservoir at site $j$.
The continuity equation then reads
\begin{equation}
\frac{d\bar n_j}{dt} =\langle J_{j-1,j}\rangle_t -\langle J_{j,j+1}\rangle_t +\langle J_{E,j}\rangle_t,
\label{eq:local_continuity}
\end{equation}
with $J_{0,1}=J_{L,L+1}=0$ for open boundary conditions. 
Equation~\eqref{eq:local_continuity} holds for both
noninteracting and interacting chains.
For the end-point geometry in Fig.~\ref{fig:fig1}, after taking the large-bias limit,
the incoming and outgoing reservoir currents are given by 
\begin{equation}
J_{\mathrm{in}}(t)=g_1[1-\bar n_1(t)], \qquad
J_{\mathrm{out}}(t)=g_L\bar n_L(t),
\label{eq:boundary_currents}
\end{equation}
where both are defined as positive in the direction
from the injecting reservoir to the extracting one.
Stationarity, i.e., the condition $d\rho^{\mathrm{ss}}/dt=0$, is realized by the NESS density matrix $\rho^{\mathrm{ss}}$ for $t \rightarrow \infty$. In the absence of additional reservoirs, this requires a spatially uniform NESS current,
\begin{equation}
J^{\mathrm{th}} = J_{\mathrm{in}}^{\mathrm{ss}} =
\langle J_{j,j+1}\rangle^{\mathrm{ss}} =
J_{\mathrm{out}}^{\mathrm{ss}},
\qquad 1\leq j<L.
\label{eq:uniform_current}
\end{equation}
If an additional weakly coupled extraction probe tip is present at site $k$, its outgoing current is given by
\begin{equation}
J_k(t)=\gamma_{\mathrm{tip}}\bar n_k(t),
\label{eq:probe_density}
\end{equation}
where $\gamma_{\mathrm{tip}}$ is the extraction rate and $\bar n_k(t)$ is evaluated for the state evolving with the
tip attached. Particle conservation  then imposes
\begin{equation}
  J_{\mathrm{in}}^{\mathrm{ss}} = J_{\mathrm{out}}^{\mathrm{ss}} + J_{k}^{\mathrm{ss}}.
\end{equation}

\subsection{Thermal benchmark}
\label{sec2c}

To distinguish structures selected by the local drive from those already favored by the Hamiltonian, we compare the NESS to  a reference TESS. To that end, we consider a global bath at temperature $T$ and chemical potential $\mu$ whose transition rates satisfy detailed balance.
The system dynamics is still described by Eq.~\eqref{eq:lindbladeq} but with the Davies jump operators \cite{Davies_1974,Davies_1976,Guo_2025}
\begin{equation}
A(\omega) = \sqrt{\gamma(\omega)} \sum_{E_m - E_n = \omega} |m\rangle\langle n| ,
\end{equation}
\noindent
where $\{|n\rangle\}$ are the many-body energy eigenstates and $E_n$ the corresponding eigenenergies of the many-body grand-canonical Hamiltonian $H - \mu N = \sum_{n} E_n \, |n\rangle\langle n|$, with $N$ total particle-number operator. Note that
 $\omega$ runs over all Bohr frequencies $\omega = E_m - E_n$, and  
 the Davies jump operators $A(\omega)$ connect energy eigenstates differing by $\omega$. The bath-dependent rates $\gamma(\omega)$ satisfy the Kubo-Martin-Schwinger condition ($\beta=1/T$),
$\frac{\gamma(+\omega)}{\gamma(-\omega)} = e^{-\beta \omega},$
which ensures thermalization to equilibrium. A common parametrization is
\begin{equation}
\gamma(\omega)=
\begin{cases}
\gamma\, n_{\rm BE}(\omega), & \omega > 0, \\[6pt]
\gamma\,\bigl[1 + n_{\rm BE}(|\omega|)\bigr], & \omega < 0,
\end{cases}
\end{equation}
with the Bose-Einstein distribution $n_{\rm BE}(\omega) = \frac{1}{e^{\beta \omega} - 1}.$
Zero-frequency terms ($\omega=0$) generate pure dephasing and must be treated separately. At zero temperature, only  transitions to lower energy states are allowed.
Note that instead of grouping all transitions with the same Bohr frequency into a single jump operator, one may resolve the dissipative dynamics at the level of individual transitions between energy eigenstates \cite{Ostilli_2017}. For each ordered pair of distinct eigenstates $|n\rangle$ and $|m\rangle$, we introduce a jump operator
$A_{mn} = \sqrt{\gamma(\omega_{mn})}\, |m\rangle\langle n|$ with $(m \neq n)$,
which induces a transition from $|n\rangle$ to $|m\rangle$, where the associated rate depends on the Bohr frequency
$\omega_{mn} = E_m - E_n$.
In the energy eigenbasis, the Davies generator implies a complete decoupling between diagonal and off-diagonal density-matrix elements. The populations
$p_n(t) = \rho_{nn}(t)$
then evolve independently of the coherences (the off-diagonal elements of $\rho$); the latter decay exponentially with time at finite $T$.
The dynamics of the populations is governed by the classical Pauli rate equation,
\begin{equation}
\frac{d p_n}{dt} = \sum_{m \neq n} \left( \Gamma_{m \to n}\, p_m
- \Gamma_{n \to m}\, p_n \right) ,
\label{eq:Pauli}
\end{equation}
where the transition rates are $\Gamma_{n \to m} = \gamma (E_m - E_n)$.
Because the rates satisfy the detailed-balance condition, the Pauli equation admits a unique (up to degeneracies eventually resolved by dephasing jump operators) stationary solution given by the Gibbs-Boltzmann distribution,
$
p_n^{\mathrm{th}} = \frac{e^{-E_n/T}}{Z},\quad Z = \sum_m e^{-E_m/T}.
$
This implies that the steady state density matrix of the full Lindblad equation is given by
\begin{equation}
\rho^{\mathrm{ss}}= \frac{e^{-(H-\mu N)/T}}{Z},
\end{equation}
which is diagonal in the energy-number ($N$) eigenbasis and represents global thermal equilibrium with chemical potential $\mu=0$. 
In the following we use this TESS as an ensemble benchmark without requiring
a specific microscopic bath realization.

\subsection{Numerical approach}\label{sec2d}

For $V=0$, the Lindblad dynamics generated by the quadratic Hamiltonian and linear jump operators closes at the level of the single-particle $L\times L$ correlation matrix $C_{ij}=\mathrm{Tr}(c_i^\dagger c_j^{}\rho)$. Writing
\begin{equation}
\mathcal H_{ij}
=-J(\delta_{i,j+1}+\delta_{i+1,j})+\mu_i\delta_{ij},
\label{eq:single_particle_H}
\end{equation}
we obtain
\begin{equation}
\dot C=i[\mathcal H^{\mathsf T},C]+\hat\Gamma
-\frac12\{\hat\Gamma+\hat\gamma,C\},
\label{eq:C-equ}
\end{equation}
where $(\hat\Gamma)_{ij}=\Gamma_i\delta_{ij}$ and $(\hat\gamma)_{ij}=\gamma_i\delta_{ij}$. 
For $V\neq0$, the correlation hierarchy does not close. The interacting results below have been
obtained by solving the many-body Lindblad equation either as a stationary linear problem with 
$\mathrm{Tr}\rho^{\mathrm{ss}}=1$ or by explicit time integration. We emphasize that no Hartree-Fock closure is needed for these results \cite{Nava_2021}. However, since the Hilbert-space dimension is $2^L$, the accessible system size is limited. Larger systems can be studied by tensor-network approaches \cite{Schollwoeck_2011,Jaschke_2019}.

Given the model and the formalism discussed above,
we now investigate how the field $F$ and the interaction $V$ will affect the stationary density profile and the particle current in a chain of length $L$. 
In particular, we compare the NESS generated by the local end-point reservoirs with a TESS reference state to check to which extent the TESS description can describe  NESS
physics. In our analysis, we study both the stationary NESS solution satisfying $\mathcal L\rho^{\mathrm{ss}}=0$, and the full time evolution of the state.

\section{Steady-state density profiles}
\label{sec3}

In this section, we study the site-resolved steady-state occupancies $\bar n_j$ for end-point reservoirs in the large-bias regime,  assuming that no mobile tip is present. 
The left environment injects particles at the left boundary ($j_L=1$), and the right environment extracts them at the right boundary ($j_R=L$). We consider a symmetric case with identical rates 
$\Gamma_1=\gamma_L=J$, where all other local rates are set to zero. The thermal TESS reference state has chemical potential $\mu=0$ and is evaluated in the zero-temperature limit. (Strictly speaking, 
the master equations used here do not apply at $T=0$ but they can be used at arbitrarily low temperatures for sufficiently low system-reservoir coupling \cite{Breuer2007}. 
It is in this sense that we consider the $T\to 0$ limit below.) 
We consider two environments (NESS vs TESS) which determine the respective stationary states in different ways. The TESS state has weights proportional to
$\exp[-\beta(E_m-\mu N_m)]$ in the many-body energy eigenbasis with total particle number $N_m$, see Sec.~\ref{sec2c}. For $T\to0$, it approaches the ground-state projector of $H-\mu N$. 
For the NESS, the local gain and loss processes instead couple eigenstates through matrix elements of $c_1^\dagger$ and $c^{}_L$.
Their energy-independent rates do not preferentially select low-energy states. Nevertheless, as shown below, these different state-selection mechanisms can produce very similar density profiles.

In the noninteracting limit ($V=0$), the distinction can be made explicit in the
single-particle eigenbasis. Let $\phi_\alpha(j)$ denote the
amplitude at site $j$ of the normalized single-particle
eigenstate $\alpha$ with energy $\epsilon_\alpha$, where
$d_\alpha$ annihilates a fermion in that eigenstate. Then
\begin{equation}
c_j=\sum_\alpha\phi_\alpha(j)d_\alpha, \quad
H=\sum_\alpha\epsilon_\alpha d_\alpha^\dagger d_\alpha.
\end{equation}
A reservoir acting locally at site $j$ therefore couples
to orbital $\alpha$ through its amplitude $\phi_\alpha(j)$
at the contact site. In the weak-contact limit of the end-point setup, when coherences between distinct orbital energies can be neglected in the population dynamics \cite{DAbbruzzo_2021},
the steady-state occupation of orbital $\alpha$,
$\nu_\alpha^{\mathrm{ss}}
=\mathrm{Tr}(d_\alpha^\dagger d_\alpha\rho^{\mathrm{ss}})$,
is given by
\begin{align}
\nu_\alpha^{\mathrm{ss}}
=\frac{a_\alpha f_1+b_\alpha f_L}{a_\alpha+b_\alpha},\quad
a_\alpha=g_1|\phi_\alpha(1)|^2,
\; \; b_\alpha=g_L|\phi_\alpha(L)|^2.
\label{eq:modal_occupation}
\end{align}
For the large-bias source-sink
configuration with $f_1=1$ and $f_L=0$, we have $g_1=\Gamma_1$ and $g_L=\gamma_L$.
The density is then approximately given by
\begin{equation}
\bar n_j\simeq\sum_\alpha|\phi_\alpha(j)|^2
\nu_\alpha^{\mathrm{ss}}.
\label{eq:modal_density}
\end{equation}
Thus an orbital coupled predominantly to the source is nearly filled, whereas one coupled predominantly to the drain is nearly empty. The stationary current follows as
\begin{equation}
J^{\mathrm{ss}} \simeq (f_1-f_L) \sum_{\alpha} 
\frac{a_{\alpha}b_{\alpha}} {a_{\alpha}+b_{\alpha}}.
\label{eq:modal_current}
\end{equation}
In Appendix~\ref{appA}, we derive these expressions and their range of validity. This approximation provides an interpretation for NESS profiles, while the noninteracting numerical results below retain the full correlation matrix of the local Lindblad equation.

Small contact amplitudes must be distinguished from exact decoupling. For a finite open chain, the single-particle
eigenvalue equation at the left end-point reads
\begin{equation}
-J\phi_\alpha(2)
=(\epsilon_\alpha-\mu_1)\phi_\alpha(1).
\end{equation}
If $\phi_\alpha(1)=0$, nonzero hopping requires
$\phi_\alpha(2)=0$. Applying the eigenvalue equation
successively at the remaining sites then forces every
component to vanish, contradicting the normalization
of the eigenstate. The same argument applies at the
right end-point. Thus every single-particle orbital has
nonzero amplitude at both end-points, although these
amplitudes can be extremely small. Localized orbitals therefore
remain coupled to the end-point reservoirs, although their relaxation can become exponentially slow, see Appendix~\ref{appA}. In particular, localization does not by itself produce dark bulk orbitals or arbitrary stationary bulk occupations in this geometry. This distinction matters when interpreting finite-time profiles at strong tilt.

\begin{figure}
\centering
\includegraphics[width=0.8\linewidth]{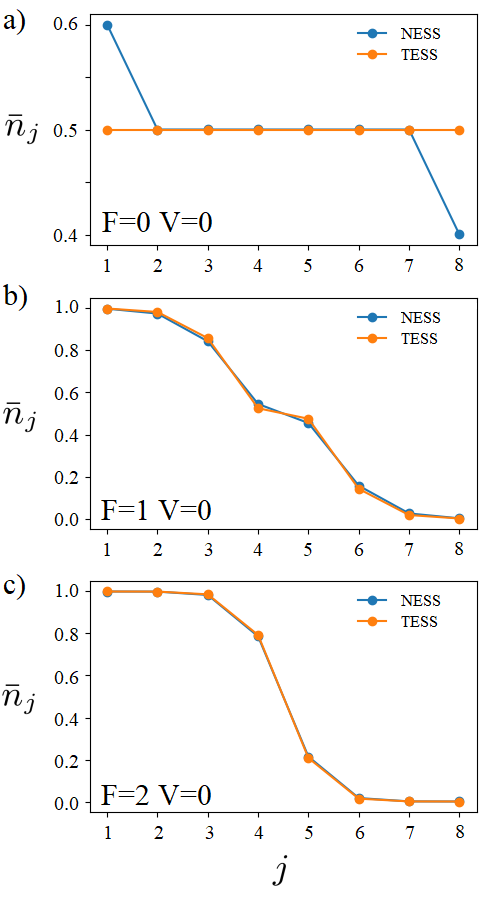}
\caption{Steady-state site occupancies $\bar n_j$ vs lattice site index $j$ for a noninteracting ($V=0$) fermion chain with $L=8$ sites and electric field parameters (a) $F=0$, (b) $F=1$, and (c) $F=2$, using units with $J=1$.  Solid curves are guides to the eye only, the actual data points are shown as dots. Note the 
difference in $\bar n_j$-scale for panel (a) with respect to panels (b,c).
Blue curves correspond to the boundary-driven NESS with $\Gamma_1=\gamma_L=1$ under large-bias conditions as obtained from the stationary solution of the Lindblad master equation \eqref{eq:lindbladeq}. Orange curves 
show the corresponding TESS reference profile with $\mu=0$ and taking the $T\to0$ limit, see Eq.~\eqref{Bessel}.}
\label{fig:fig2}
\end{figure}

Similarly, the TESS density profile at $V=0$ follows from the eigenenergies $\epsilon_\alpha$ and eigenstates $\phi_\alpha$,
\begin{equation}
\bar n_j^{\mathrm{th}}
=\sum_\alpha|\phi_\alpha(j)|^2
\frac{1}{1+e^{\beta(\epsilon_\alpha-\mu)}}.
\label{eq:finite_density}
\end{equation}
This result is exact and holds for finite chains.
An analytical expression valid for finite $F$ and in the $T \rightarrow 0$ limit with $\mu=0$ can be obtained from the infinite-chain Wannier-Stark orbitals, 
\begin{equation}\label{Bessel}
    \bar{n}^{\rm th}_j \simeq 1 - \sum_{m=L/2 +1}^{L} \mathbf{J}^2_{j-m}(x) 
    \simeq \sum_{m=1}^{L/2} \mathbf{J}^2_{j-m}(x) ,
\end{equation}
where $\mathbf{J}_{\alpha}(x)$ is the cylindrical Bessel function of order $\alpha$ and $x=2J/F$. For simplicity, we have here assumed the case of even (but large) $L$. Both expressions quoted in Eq.~\eqref{Bessel} become exact and equivalent for $L \rightarrow \infty$. For finite $L$, the first expression is a better approximation for $1 \leq j \leq L/2$, i.e., for the left half of the chain, while the second expression is preferable for $L/2<j \leq L$.  
Figure~\ref{fig:fig2} compares the noninteracting NESS and TESS profiles for different field parameters $F$. First, for $F=0$, see Fig.~\ref{fig:fig2}(a), the TESS gives a uniformly half-filled profile. The NESS also has $\bar n_j=1/2$ in the interior of the chain but the end-point occupancies are $\bar n_1=0.6$ and $\bar n_L=0.4$ for the parameters shown. These boundary deviations sustain a finite current and distinguish the NESS state from the equilibrium state 
even if the bulk densities agree. Next, for $F=1$, see Fig.~\ref{fig:fig2}(b), both profiles become strongly polarized, with particles concentrated on the left side of the chain which has a lower potential. Their shapes
are close, with small differences in the central region of the chain. Finally, for $F=2$, see Fig.~\ref{fig:fig2}(c), the polarization effect sharpens further and both profiles now basically coincide.

In the NESS, the field localizes the orbitals and makes their coupling to the two contacts strongly asymmetric. Orbitals centered on the left are predominantly populated by the source, whereas orbitals centered on the right are predominantly depleted by the drain. Both mechanisms consequently favor an almost filled left region and an almost empty right region, separated by a density crossover. In the TESS, this polarization follows from filling the low-energy orbitals. Agreement of these one-point observables thus does not imply a matching of both density matrices. In fact, other observables, for example, the steady-state current, can readily detect the difference.

\begin{figure}
\centering
\includegraphics[width=0.8\linewidth]{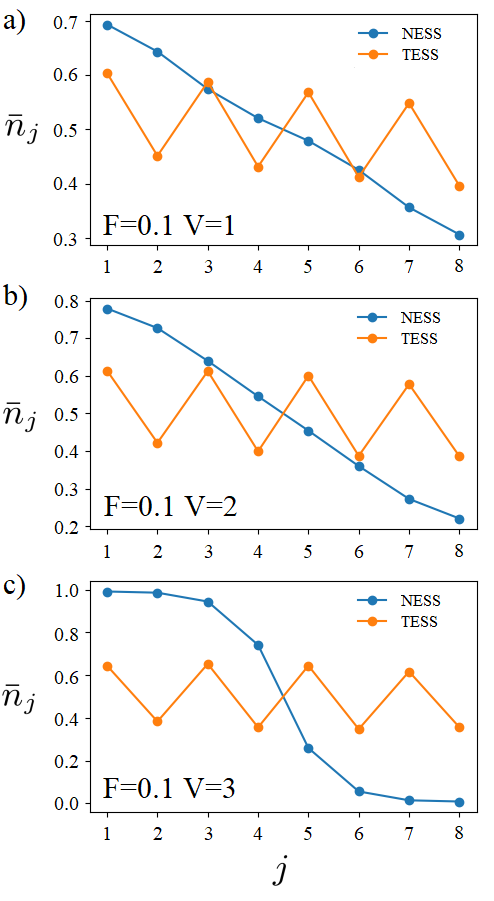}
\caption{Steady-state site occupancies $\bar n_j$ vs $j$ for a chain with $L=8$ sites with a weak tilt
$F=0.1$, with interaction parameter (a) $V=1$, (b) $V=2$, and (c) $V=3$.  We again put $J=1$ and consider large-bias conditions for the NESS as in Fig.~\ref{fig:fig2}. 
Note the differences in the $\bar n_j$-scale in the different panels.
Blue curves are for the boundary-driven NESS, and orange curves for the TESS.  All other parameters are as in Fig.~\ref{fig:fig2}.}
\label{fig:fig3}
\end{figure}

Interactions produce a clearer distinction between the NESS and
TESS at weak tilt, as shown in Fig.~\ref{fig:fig3} for $F/J=0.1$. For all three values of $V$ studied  in Fig.~\ref{fig:fig3}, the TESS profile alternates between large 
odd-site and small  even-site occupancies. Repulsive interactions favor alternating configurations, while hopping reduces their density contrast. The small positive field biases their orientation in this finite open chain. For even $L$ and $F=0$, the Hamiltonian is particle-hole (and even-odd) symmetric and the ground state of the closed system would show a flat occupancy. For finite $F$, we refer to the observed modulation as CDW-like.
The NESS profiles in Fig.~\ref{fig:fig3} show a rather different
structure. For $V/J=1$ and $V/J=2$, see Figs.~\ref{fig:fig3}(a,b), the density $\bar n_j$ decreases smoothly as one moves from $j=1$ to $j=L$, with a larger end-point imbalance at $V/J=2$. 
For $V/J=3$, see Fig.~\ref{fig:fig3}(c), particles accumulate on the left side while the right side becomes almost empty, producing a pronounced domain wall behavior. In this regime, the large-bias drive
favors a polarized configuration similar to the one observed in the noninteracting case, rather than the CDW-type low-energy configurations predicted by the TESS reference state.
The end-point densities connect this comparison directly to
transport. In fact, Eq.~\eqref{eq:boundary_currents} gives
\begin{equation}
J^{\mathrm{ss}} =\Gamma_1(1-\bar n_1)=\gamma_L\bar n_L.
\label{eq:density_endpoint_current}
\end{equation}
The nearly filled left end site ($j=1$) and the nearly empty right end site ($j=L$) observed for the NESS in
Fig.~\ref{fig:fig3}(c) therefore imply a strongly suppressed current. This behavior is consistent with interaction-induced blocking domains 
as previously found in strongly driven fermion chains \cite{Mendoza_2024}.

\begin{figure}
\centering
\includegraphics[width=0.8\linewidth]{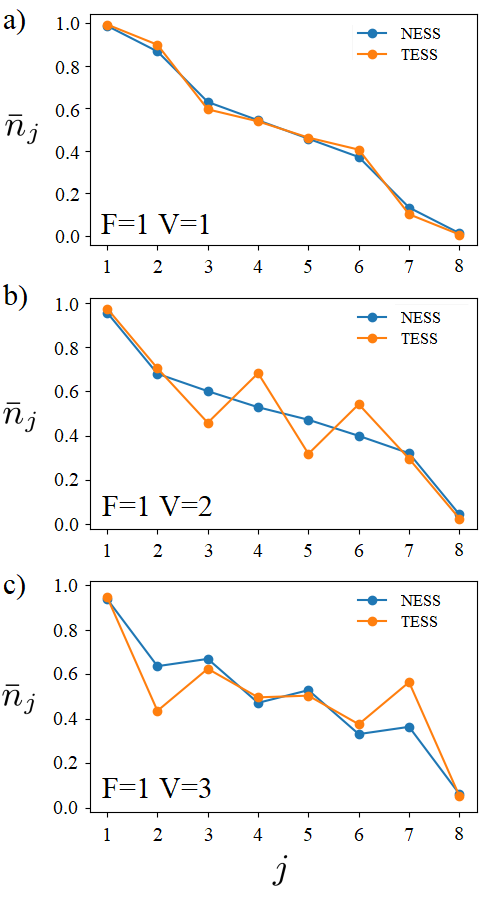}
\caption{Steady-state site occupancies $\bar n_j$ vs $j$ for $L=8$ and $F=1$
with (a) $V=1$, (b) $V=2$, and (c) $V=3$, using $J=1$.
Blue curves are for the boundary-driven NESS under strong-bias conditions. Orange curves describe 
the TESS reference state. Other parameters are as in Fig.~\ref{fig:fig2}. }
\label{fig:fig4}
\end{figure}

Next we consider a stronger tilt, $F/J=1$, where we find that the interplay between polarization and CDW-type staggering effects strongly depends on the interaction parameter $V$, see Fig.~\ref{fig:fig4}. 
For $V/J=1$, see Fig.~\ref{fig:fig4}(a), the NESS and TESS density profiles are quite close and tend to decrease with increasing  site index $j$. For $V/J=2$, see Fig.~\ref{fig:fig4}(b), the TESS develops CDW-type
oscillations in the central region, while the NESS profile is monotonically decreasing. For $V/J=3$, see Fig.~\ref{fig:fig4}(c), both profiles are non-monotonic, but the respective oscillations differ substantially in amplitude and spatial distribution. 
The NESS profile exhibits weak alternating variations superimposed on an overall density gradient, whereas the oscillations are more pronounced for the TESS reference solution, in particular near sites $j=2$ and $j=7$. 
Compared to Fig.~\ref{fig:fig4}(a), the occupation of site $j=1$ in the NESS decreases and the 
occupation of site $j=L$ increases. Equation~\eqref{eq:density_endpoint_current} then implies that increasing the interaction from $V/J=1$
to $3$ will enhance the steady-state current for this value of $F$.
This enhancement illustrates that the effects of tilt ($F$) and interaction ($V$) on transport are not simply additive. In the
present geometry, particles are injected at the low-potential end and extracted at the high-potential end. This corresponds
to the forward-driving configuration of Ref.~\cite{Mendoza_2024}, where competition between
interaction and tilt produces conduction resonances associated with avoided crossings of many-body energy levels.
Our density profiles are consistent with interaction-assisted transport, although identifying individual resonances requires
examining the current and the many-body spectrum. 

\begin{figure}
\centering
\includegraphics[width=0.8\linewidth]{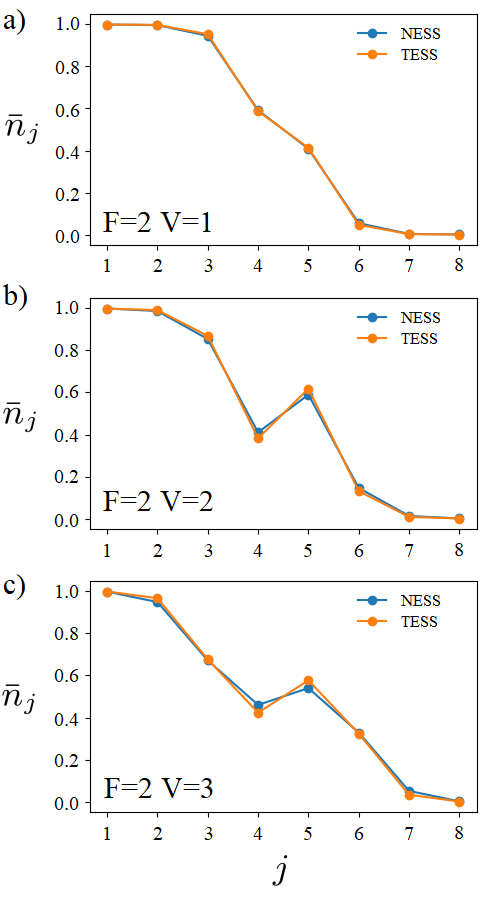}
\caption{Steady-state site occupancies $\bar n_j$ vs $j$ for $L=8$ and $F=2$,
with (a) $V=1$, (b) $V=2$, and (c) $V=3$, using $J=1$.
Blue curves are for the boundary-driven NESS under strong-bias conditions. Orange curves
for the TESS. Other parameters are as in Fig.~\ref{fig:fig2}.}
\label{fig:fig5}
\end{figure}

At the even stronger tilt $F/J=2$, the NESS and TESS profiles nearly
coincide for all three interaction strengths studied in
Fig.~\ref{fig:fig5}. For $V/J=1$, see Fig.~\ref{fig:fig5}(a), the density profile decreases
monotonically, with almost filled vs empty outer regions and a crossover region near the center of the chain. For $V/J=2$, see Fig.~\ref{fig:fig5}(b),
both states exhibit a central inversion, $\bar n_4<\bar n_5$,
while the density decreases otherwise when moving along the chain. Finally, for $V/J=3$, see Fig.~\ref{fig:fig5}(c),
this inversion persists with smaller contrast, while the crossover region extends deeper into the bulk of the chain. Increasing
$V/J$ therefore changes both the width and the internal structure
of the crossover region, rather than simply increasing its CDW-type staggered modulation. 
The close agreement between the NESS and TESS profiles at $F/J=2$ contrasts with their marked differences for $F/J=0.1$ observed in Fig.~\ref{fig:fig3}.

In equilibrium, these trends reflect competition between the electric field parameter $F$, which favors spatial polarization, and the electron-electron repulsion parameter $V$, which
favors alternating occupations \cite{Boidi_2025}. A local energy argument also helps to interpret the driven system.
For a rightward hop on an interior bond $(j,j+1)$, the change in diagonal energy is
\begin{equation}
\Delta E=F+V(n_{j+2}-n_{j-1}).
\label{eq:hop_detuning_main}
\end{equation}
For $n_{j-1}=1$ and $n_{j+2}=0$, the interaction lowers the
hopping cost from $F$ to $F-V$. It can therefore facilitate
the rearrangement of a polarized configuration when the
interaction and field energies compensate. 
The derivation and its relation to many-body avoided crossings are discussed in Appendix~\ref{appA}.

A complementary consideration is that the interaction strength $V$ does not directly measure how efficiently particles move between Wannier-Stark orbitals. Part of the interaction only changes the energies of their
occupation configurations, while other terms transfer particles
between orbitals. At strong tilt, the orbitals become concentrated on individual sites, and these transfer terms become small at fixed $V$ \cite{Krajewski_2025}. Interactions can therefore substantially affect configuration energies while producing only slow redistribution of particles.

Similar TESS and NESS density profiles are
expected when the field dominates both hopping and
interactions. In this regime, the low-temperature thermal
state favors an occupied left region and an empty right
region, consistent with the orientation imposed by the
source and drain. Agreement persists when interactions
modify the central region, as in Fig.~\ref{fig:fig5}, but
is then not guaranteed by the relative magnitudes of $F$
and $V$.  We note that symmetry relations constrain the TESS and NESS density profiles, in particular their behavior  
under a reversal of the directions of  $F$ and/or $V$, or when exchanging the left and right reservoirs (backward-driving regime).  We discuss these symmetry relations in
Appendix~\ref{symmetries}.

Finally, we emphasize that the close agreement between the NESS and TESS
density profiles observed in several cases above does not imply thermalization of the driven
state. In fact, the stationary current also depends on coherences
between many-body eigenstates, which are not determined by the site occupancies. 
The current considered in Sec.~\ref{sec4} therefore provides complementary information.  

\section{Nonequilibrium dynamics}
\label{sec4}

In this section, we turn to the steady-state current and also address the full nonequilibrium dynamics of the state as it approaches the NESS.   
In Sec.~\ref{sec4a}, we analyze these questions in the absence of the mobile tip.
We discuss a penetration length over which currents can effectively be transmitted in Sec.~\ref{sec4b}, 
again without a mobile tip.
Finally, we discuss time-dependent profiles and the mobile-tip setup for measuring the density profile through currents flowing toward the extraction reservoir in Sec.~\ref{sec4c}.

\subsection{Steady-state currents and time evolution}\label{sec4a}

We here examine the time-dependent approach of $\rho(t)$ to the NESS, $\rho^{\rm ss}=\rho(\infty)$, and the associated dependence
of the stationary current $J^{\rm ss}$ on $F/J$ and $V/J$ in the
end-point setup without the mobile tip. As initial condition imposed at time $t=0$, we consider an empty chain,
$\rho(0)=|0\dots0\rangle\langle0\dots 0|$.  We again study the large-bias regime as defined in Sec.~\ref{sec2b}, where the incoming and outgoing currents
satisfy Eq.~\eqref{eq:boundary_currents} and their difference determines the rate of change, $dN/dt$, of the
total particle number $N(t)=\sum_j \bar n_j(t)$.
At $t=0$, the incoming particle current at site $j=1$ therefore equals $\Gamma_1$,
while the current outgoing at site $j=L$ vanishes. As particles
populate the chain, the occupation of site $j=1$ reduces
 injection by the left reservoir through Pauli blocking, while the occupation of
 site $j=L$ enables extraction by the right reservoir \cite{Nava_2021}.

\begin{figure}
\centering
\includegraphics[width=0.75\linewidth]{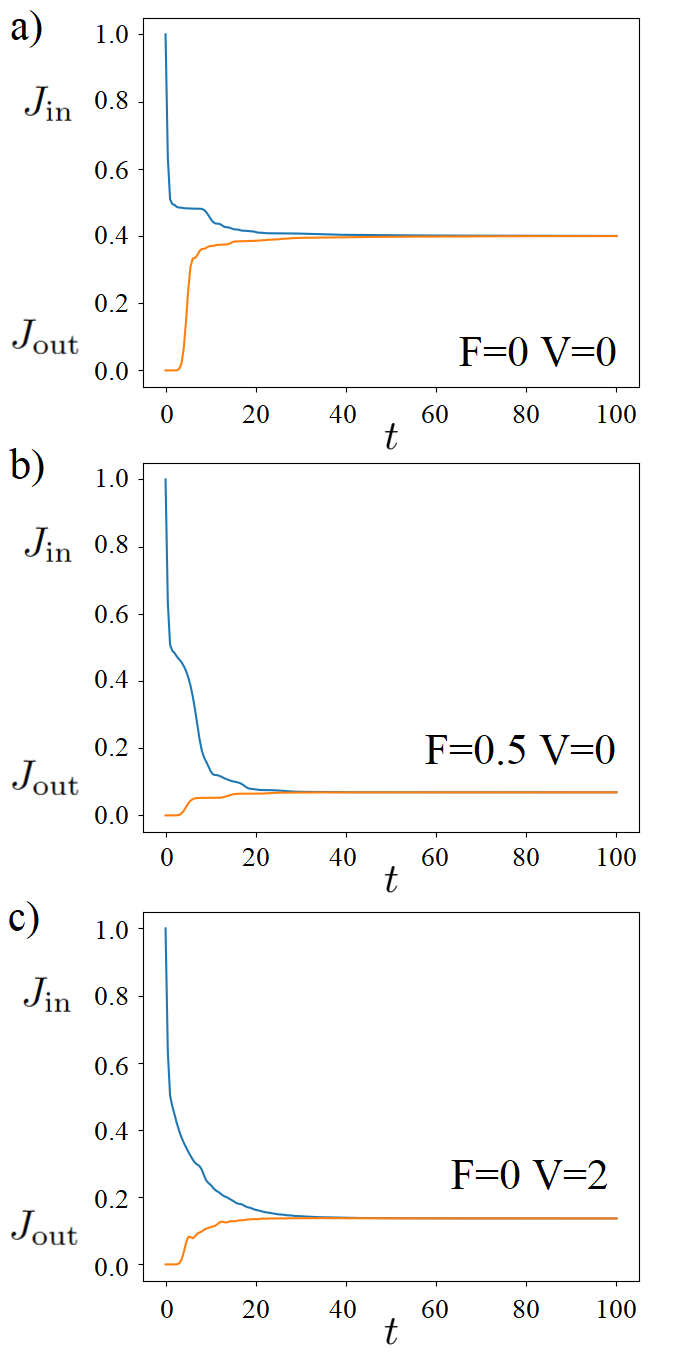}
\caption{Time evolution of the incoming current
$J_{\mathrm{in}}(t)$ (blue) and outgoing current
$J_{\mathrm{out}}(t)$ (orange), starting from an
empty chain at $t=0$. The parameters are $L=8$, $J=1$,
$\Gamma_1=\gamma_L=1$, with (a) $F=V=0$,
(b) $F=0.5$ and $V=0$, and (c) $F=0$ and $V=2$.}
\label{fig:fig6}
\end{figure}

Figure~\ref{fig:fig6}(a) shows the dynamics in the
homogeneous ($F=0$) noninteracting ($V=0$) chain. The incoming current
first decreases rapidly and develops a short plateau,
while the outgoing current remains zero until particles
have propagated across the system. The dispersion
$\epsilon(k)=-2J\cos k$ gives a maximum group velocity
$v_{\max}=2J$, with lattice spacing set to unity.
The corresponding traversal-time estimate is
\begin{equation}
t_{\mathrm{tr}}\simeq\frac{L-1}{2J}.
\label{eq:traversal_time}
\end{equation}
For $L=8$ and $J=1$, this gives $t_{\mathrm{tr}}\simeq3.5$,
consistent with the onset of appreciable extraction.
This is a propagation time scale rather than a sharp
arrival time.

The early plateau describes a transient regime before
the distant boundary appreciably affects injection.
Subsequent propagation and partial reflection at the
contacts produce the small steps visible during relaxation.
The currents eventually approach the same steady-state value,
$J^{\mathrm{ss}}=0.4$ for the parameters shown.
The early plateau should therefore be distinguished
from the stationary current of the finite chain  where Eq.~\eqref{eq:density_endpoint_current} is realized.
A finite tilt substantially changes this behavior, see
Fig.~\ref{fig:fig6}(b). For $F=0.5$ and $V=0$, the
incoming current drops well below the initial transient
value, and the stationary current is approximately
$0.07$. The field suppresses transmission between the
contacts, consistent with the orbital-coupling picture
discussed above. The early-time outgoing current remains small,
while particles accumulate near the injecting boundary
and reduce further injection.

Interactions also suppress the current in the untilted
chain, see Fig.~\ref{fig:fig6}(c) for $F=0$ and $V=2$,
where the incoming current decreases over a broader transient
while the outgoing current rises toward the common
stationary value $J^{\rm ss}\approx 0.14$.
Thus both tilt and repulsion reduce the transmitted
current relative to the homogeneous noninteracting
reference, although their transient signatures differ.
In all three panels, the boundary currents become
indistinguishable on the plotted scale at sufficiently
long times. Their agreement is a useful particle-balance
check, but it establishes only that the total occupation
has stopped changing. It does not by itself imply
that every local density, correlation, or matrix
element of $\rho(t)$ has reached its stationary value.
In particular, slow redistribution within the chain
can have little effect on the end-point occupancies.
Convergence of the boundary currents must therefore
be distinguished from convergence of the full state.

\begin{figure}
\centering
\includegraphics[width=0.95\linewidth]{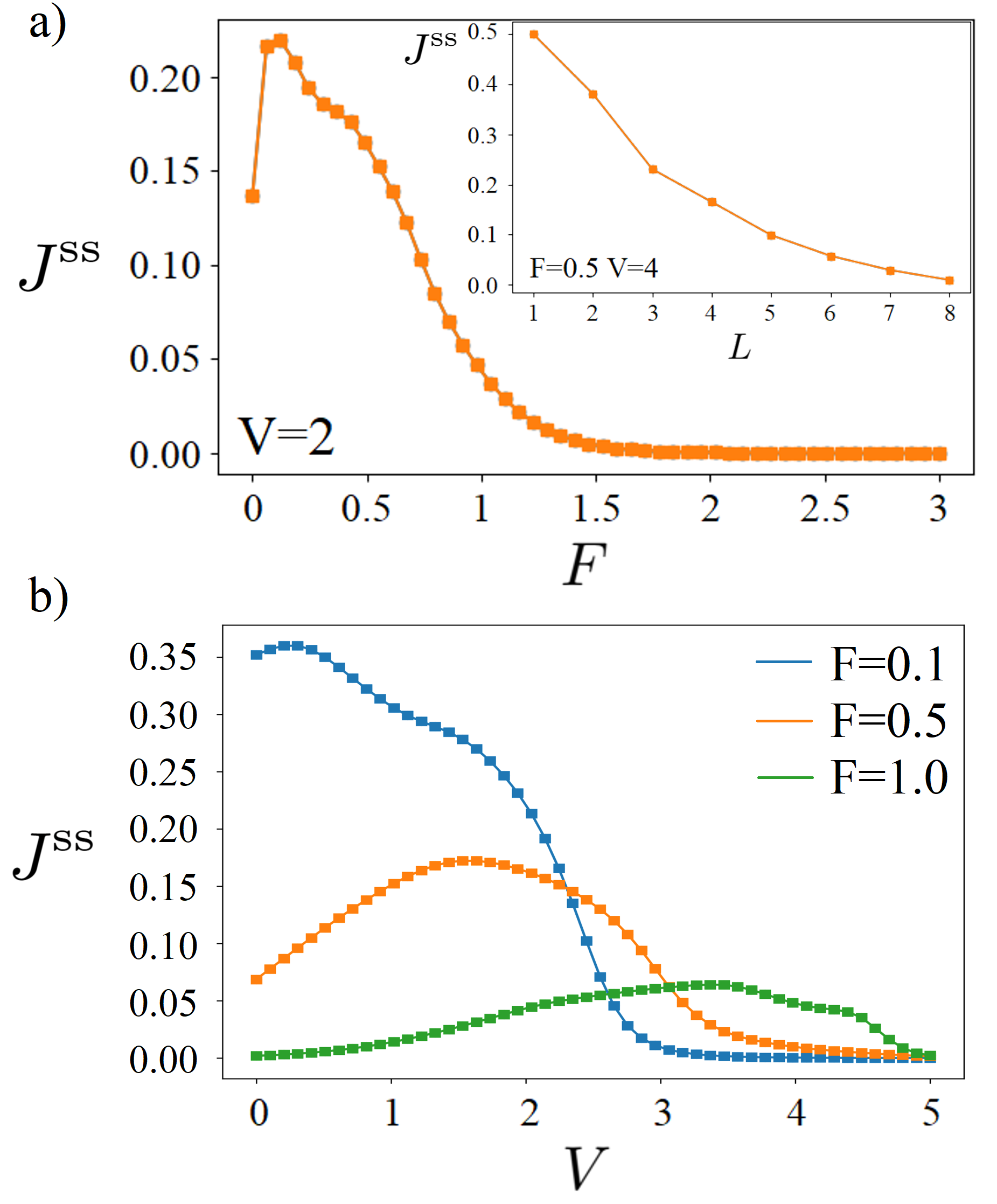}
\caption{Stationary current $J^{\mathrm{ss}}$ for $L=8$ and
$\Gamma_1=\gamma_L=J=1$, where we show 
(a) the dependence on $F$ at fixed $V=2$, and
(b) the dependence on $V$ at fixed $F=0.1$ (blue),
$F=0.5$ (orange), and $F=1$ (green).
The inset in panel (a) shows the dependence of $J^{\rm ss}$ on the chain length $L$
for $F=0.5$ and $V=4$.}
\label{fig:fig7}
\end{figure}

Figure~\ref{fig:fig7} shows that the effects of   $F$ and   $V$ on the stationary current $J^{\rm ss}$ can be
nonmonotonic. At fixed $V/J=2$, see Fig.~\ref{fig:fig7}(a), a weak tilt
initially increases the current above its value at
$F=0$. A maximum then occurs near $F/J\simeq0.1$, followed
by an overall decrease as the field $F$ grows.
Thus adding a field to the interacting chain can
initially facilitate transport, even though a
sufficiently strong field suppresses it.
The complementary scans in Fig.~\ref{fig:fig7}(b)
show how the interaction dependence changes with
the tilt. At $F/J=0.1$, the current has a shallow
maximum at weak interaction and then decreases,
with a pronounced drop between (approximately)
$V/J=2$ and $3$. At $F/J=0.5$, the $V=0$ current becomes smaller than for $F/J=0.1$, and
the current maximum shifts to the larger interaction strength $V/J\simeq1.5$. Finally, for $F/J=1$,
the noninteracting current is strongly
suppressed. However, increasing $V/J$ restores an appreciable
current, with a broad maximum around $V/J\simeq3.3$.
For even larger $V/J$, the current decreases again.
Increasing the tilt parameter $F$ therefore
moves the current maximum toward stronger
interactions while reducing its height.

These observations are consistent with a competition
between the field energy and the interaction energy.
As discussed in Appendix~\ref{appA},
an interaction-energy reduction can compensate the
cost of moving a particle uphill. In the full
many-body system, conduction enhancements depend
on the associated state mixing, reservoir populations,
and current-carrying coherences. This mechanism is
related to the forward-driving resonances identified
in Ref.~\cite{Mendoza_2024}.
The maxima in Fig.~\ref{fig:fig7} need not occur at
$F=V$ and in general depend on the full many-body spectrum.

The current becomes very small when either the tilt
or the interaction dominates. This suppression mechanism is consistent with the
formation of polarized, weakly transmitting configurations. At small tilt $F/J\ll 1$, the system acts as driven interacting conductor in a ballistic transport regime for $V/J\ll 1$.
However, for strong repulsion,  the large-bias drive can favor polarized, weakly transmitting configurations. The finite-chain results establish current suppression. For stronger tilting, $F/J\gg 1$, the Stark blockade takes place.
In particular, for $V=0$, the coherent hopping is suppressed with increasing $F$ and the system becomes insulating due to Wannier-Stark localization. For moderate  $V/J$, 
interactions can restore transport through the interaction-enabled hopping mechanism. Further increasing $V/J$ then strongly constrains the particle motion and the current is suppressed.
The inset of Fig.~\ref{fig:fig7}(a) provides a complementary spatial test for this scenario. Using $F/J=0.5$ and $V/J=4$, we here study the $L$-dependence of the steady-state current under strong-bias conditions.
Clearly, the current decreases as the chain becomes longer. This observation implies that one can use transport for different contact separations (or system sizes) to determine the distance over which particles are effectively transmitted, i.e., the penetration length.  We examine this length scale next.

\subsection{Penetration length from length-dependent transport}
\label{sec4b}

The suppression of the stationary current with increasing
source-drain separation provides a way to characterize
the distance over which particles can be transmitted through
the tilted chain. We implement this idea by considering a family 
of chains with different $L$,
each coupled to an injecting reservoir at site $j=1$ and an
extracting reservoir at site $j=L$ under symmetric large-bias conditions.  Again, we assume that no mobile tip is present. The field ($F$), interaction ($V$),
and reservoir rate $(g)$ are kept fixed throughout the scan, where we put $J=1$.
For each $L$, we determine the stationary current
$J^{\mathrm{ss}}(L)$ and compare it with a
specified small detection threshold $J^{\mathrm{th}}$. We first define the largest integer value of $L$ for which the current exceeds this threshold,
\begin{equation}
L^\ast(F,V,g,J^{\mathrm{th}})=\max\{L:\ J^{\mathrm{ss}}(L)>J^{\mathrm{th}}\}.
\label{eq:Lstar_def}
\end{equation}
The \emph{penetration length} is then defined as 
\begin{equation}\label{eq:penle}
    d^\ast=L^\ast-1.
\end{equation}
Physically, $d^\ast$ corresponds to the separation length between the left and right reservoir in units of the lattice spacing such that $J^{\rm ss}(d^\ast+2)<J^{\rm th}<J^{\rm ss}(d^\ast+1)$. For $d^\ast=0$, the current for a chain with $L=2$ sites is thus already below the threshold value.

Varying the total length $L$ of the chain is numerically advantageous,
particularly for the interacting problem  whose
Hilbert-space dimension grows as $2^L$. It allows us
to vary the distance traversed by   particles while
simulating only the segment between the reservoirs.
Physically, the calculation measures the transmission
of an isolated segment with fixed end-point couplings.
In a localized regime, a similar suppression with increasing
separation is expected for reservoirs attached inside
a longer chain if the exterior sites only weakly
perturb the states responsible for transmission. 
However, slow relaxation, quantum coherence, and/or additional interaction pathways can make this hypothesis questionable,
and our direct procedure without exterior sites avoids such complications from the outset.

A simple energetic argument provides an estimate
for the transport crossover. For $V=0$,
the hopping dispersion $\epsilon(k)=-2J\cos k$ allows
kinetic energies between $-2J$ and $2J$. Now add the potential
$\mu_j=F(j-j_c)$. In a local-band
picture, a particle of energy $E$ can therefore propagate
where $|E-\mu_j|\leq2J$. Here $E-\mu_j$ is its kinetic energy. As the particle moves uphill, $\mu_j$ increases by $F$ per site, so its kinetic energy decreases. It can compensate a potential increase of at most $4J$ before leaving this locally allowed range. The right and left end-points overlap only if $\mu_L-\mu_1 = F(L-1) \le 4J$.
Once the contact separation exceeds this scale,
transmission relies increasingly on the small tails
of localized orbitals. For the penetration length, this argument yields the estimate
\begin{equation} \label{eq:V0_penetration_length_estimate}
d_{\mathrm{est}}=L_{\mathrm{est}}-1\sim\frac{4J}{F}.
\end{equation}
It identifies a transport crossover rather than
a sharp cutoff or the precise threshold length $d^\ast$.
The same interpretation follows from the orbital
description in Eq.~\eqref{eq:modal_current}. An orbital
contributes to transmission only if it overlaps with both
reservoir sites. Increasing their separation suppresses
these simultaneous overlaps and consequently reduces
the current. Interactions modify the energies and
mixing of the configurations involved in transport.
As discussed above, they can compensate part of the
field cost and enhance transmission, but they can also
produce blocking configurations. Their effect on
the penetration length is therefore determined
numerically.

\begin{figure}
\centering
\includegraphics[width=0.75\linewidth]{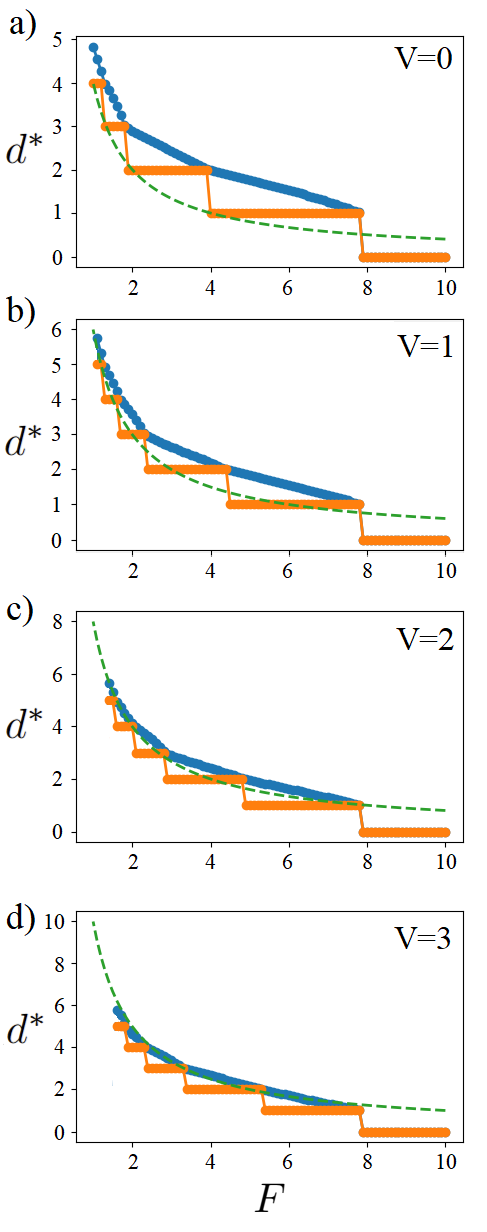}
\caption{Penetration length $d^\ast$ (in units of the lattice spacing) vs field parameter $F$ (in units of $J$) obtained 
by varying the total length of a chain with end-point reservoirs
using large-bias conditions with $\Gamma_1=\gamma_L=J=1$ and 
current threshold $J^{\mathrm{th}}=0.03$. Results are shown for interaction parameters
(a) $V=0$, (b) $V=1$, (c) $V=2$, and (d) $V=3$.
Orange symbols show numerical results obtained from  Eqs.~\eqref{eq:Lstar_def} and \eqref{eq:penle}. The plateau structure
arises because $d^\ast$ has to be integer. Blue symbols show
a continuous estimate obtained by interpolating
the length-dependent stationary current.
Green dashed curves show the equilibrium
scale $\xi_{1,{\rm int}}$ in Eq.~\eqref{eq:xi1} with $c=0$.}
\label{fig:fig8}
\end{figure}

Figure~\ref{fig:fig8} shows the resulting penetration
length for a specific choice of $J^{\mathrm{th}}$. The discrete estimate
(orange) changes in steps because the chain length
is integer. We   obtain a continuous estimate
(blue) by interpolating the stationary current
between neighboring lengths that bracket the
threshold.
If the current is nonmonotonic in $L$, the largest
crossing must be selected, consistent  with
Eqs.~\eqref{eq:Lstar_def} and \eqref{eq:penle}.
For $V=0$, see Fig.~\ref{fig:fig8}(a), the penetration
length $d^\ast$ decreases with increasing $F$, consistent
with the confinement of Wannier-Stark orbitals
and the estimate in Eq.~\eqref{eq:V0_penetration_length_estimate}. We note that $d^\ast$  shifts upward (downward) upon decreasing (increasing)   $J^{\rm th}$.
The discrete curve passes through successive integer plateaus, while the interpolated curve
resolves the overall decrease.

The curves shown in Figs.~\ref{fig:fig8}(b)--(d) show a similar
overall decrease of $d^\ast$ with $F$ for interacting cases, where the transitions
between the shortest-length plateaus move toward
larger fields with increasing $V$. In particular,
the drop from $d^\ast=2$ to $d^\ast=1$ occurs
at progressively stronger tilt. Over this
parameter range, interactions thus allow
detectable transmission to survive against
a larger potential gradient. This is again consistent
with the interaction-assisted transport
discussed above.

A useful comparison is provided by the localization
length estimated in Ref.~\cite{Boidi_2025}
from the TESS density profile where a
noninteracting argument identifies the spatial
interval over which the local potential crosses
the bandwidth, giving the same scale
$4J/F$ as Eq.~\eqref{eq:V0_penetration_length_estimate} for the penetration length.
For finite repulsion, $V>0$, the estimated boundaries
of the intermediate-density region are
$\mu_j=\pm(2J+V)$, leading to the estimate 
\begin{equation}
\xi_{1,\mathrm{int}}
=\frac{4J+2V}{F}+c,
\label{eq:xi1}
\end{equation}
where a small additive constant $c$ may arise due to lattice-scale
offsets. Ref.~\cite{Boidi_2025}, taking into account Eq.~\eqref{eq:penle}, predicts $c=0$.
The quantity $\xi_{1,\mathrm{int}}$ estimates the total equilibrium crossover width
separating nearly filled and nearly empty regions, defined in analogy to the penetration length $d^\ast$ in Eq.~\eqref{eq:penle}.
If a CDW develops, this width includes both the central modulated region and its crossovers
to the outer regions. 

The close matching between the equilibrium scale $\xi_{1,{\rm int}}$ and the numerically determined transport penetration length in the NESS, see Fig.~\ref{fig:fig8}, has a natural
interpretation. In the noninteracting case, the orbitals that determine the density crossover also mediate transmission between the reservoirs.
Both observables therefore reflect the same spatial confinement. For the interacting system, 
the comparison suggests that the enlarged energy window $4J+2V$ also controls the
transmission range of the nonequilibrium system for the studied parameter range. Note that a quantitative matching is not expected since $d^\ast$
additionally depends on the reservoir coupling strengths and on the current threshold.
We conclude that a length scan of the current can provide an operational measure of spatial transmission properties as encoded by the penetration length, without requiring knowledge of the site-resolved density profile.
 
\subsection{Scanning staggered density profiles}
\label{sec4c}

At equilibrium, repulsive interactions favor alternating occupations in the
corresponding closed tilted chain, where a central CDW region
can develop whose extent increases with $V$ and decreases
with $F$ \cite{Boidi_2025}. The boundary-driven system also
exhibits staggered density profiles for some parameters,
as illustrated in Figs.~\ref{fig:fig4}(c) and
\ref{fig:fig5}(b). We now examine how these profiles develop from an initially empty chain and how a weak mobile
extraction reservoir, see Fig.~\ref{fig:fig1} for a schematic sketch, can be employed in experiments to resolve their spatial structure.

\begin{figure}
\centering
\includegraphics[width=0.75\linewidth]{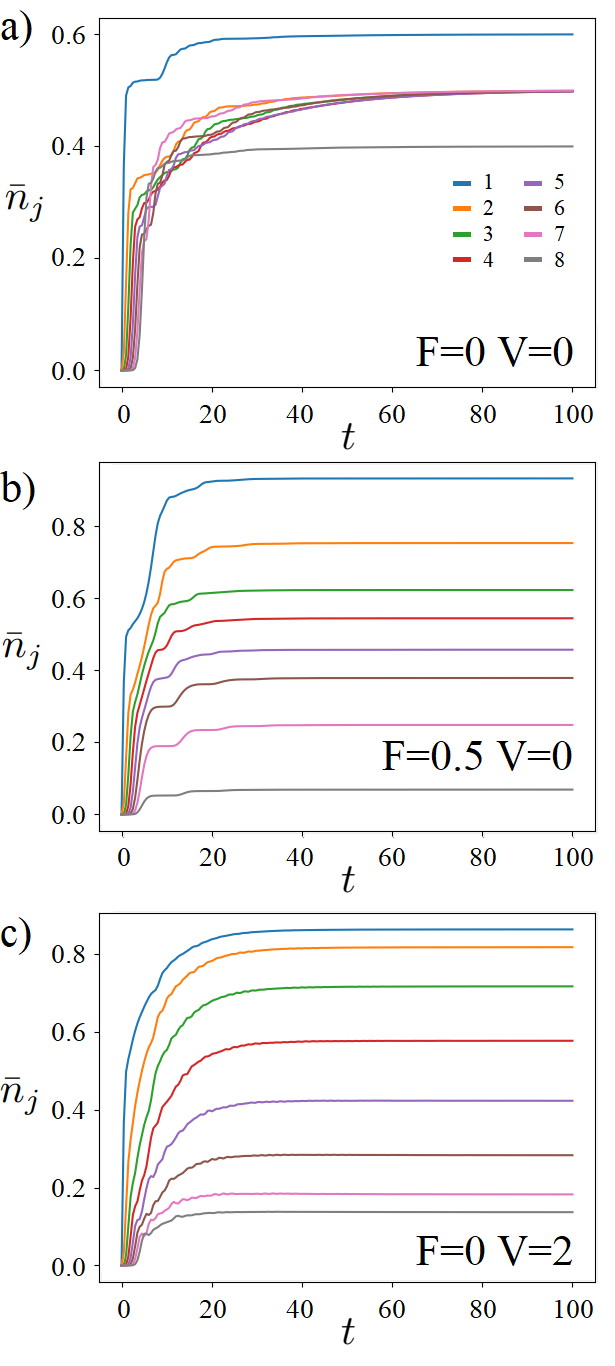}
\caption{Time evolution of the site occupancies $\bar n_j(t)$ from an initially empty chain in the
end-point reservoir setup for $L=8$ and $\Gamma_1=\gamma_L=J=1$. Results are shown for
(a) $F=V=0$, (b) $F=0.5$ and $V=0$, and (c) $F=0$ and $V=2$.
Each curve corresponds to a different site, see the legend in panel (a).}
\label{fig:fig9}
\end{figure}

Figure~\ref{fig:fig9} illustrates the filling dynamics
for three parameter sets without a staggered stationary
density profile. Particles enter through the left
boundary and propagate into the chain, so an appreciable
occupation develops progressively at more distant sites.
Coherent propagation and reflections produce transient
oscillations and step-like features, similar to those in Fig.~\ref{fig:fig6} for the incoming and outgoing currents. 

For the clean noninteracting case,  $F=V=0$, see Fig.~\ref{fig:fig9}(a), the bulk
occupancies approach $1/2$, while the first and last  sites approach $\bar n_1=0.6$ and $\bar n_L=0.4$, respectively, as expected for equal-strength reservoirs, see Fig.~\ref{fig:fig2}(a).  We note that Ref.~\cite{Nava_2021} provides corresponding analytical expressions of the occupancies and currents.
A finite field $F$, see Fig.~\ref{fig:fig9}(b), causes a polarized stationary profile, with larger occupations near the left chain end. Interactions
at zero field, see Fig.~\ref{fig:fig9}(c), also produce an appreciable density gradient under the imposed large-bias drive.
We emphasize that relaxation of bulk densities need not occur on the same time scale as relaxation of the boundary
currents. In particular, nearly stationary boundary occupancies can coexist with ongoing redistribution processes inside the chain. 
Equality of the incoming and outgoing currents implies that the total particle number has stopped changing at that point, but this does not yet establish
stationarity of every local density. A spatial scan must therefore be extended to sufficiently long times to allow all bulk occupations to relax.

\begin{figure}
\centering
\includegraphics[width=0.75\linewidth]{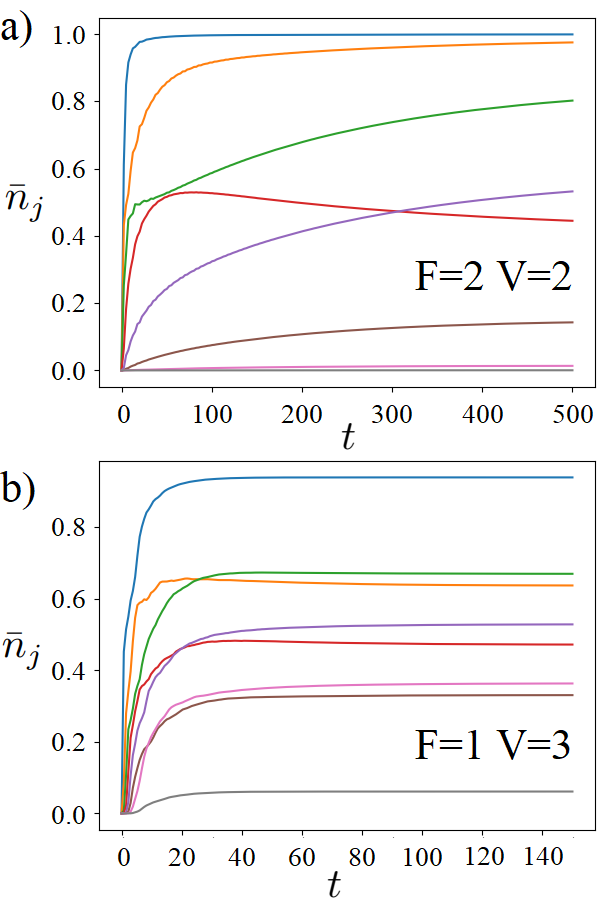}
\caption{Time evolution of the site occupancies from an initially empty chain for parameters associated with
staggered NESS profiles for $L=8$ and $\Gamma_1=\gamma_L=J=1$.  Different colors correspond to different sites as shown in the legend of Fig.~\ref{fig:fig9}(a). 
Results are shown for (a) $F=V=2$ and (b) $F=1$ and $V=3$.  The corresponding stationary density profiles are shown in
Figs.~\ref{fig:fig5}(b) and \ref{fig:fig4}(c),
respectively.}
\label{fig:fig10}
\end{figure}

Figure~\ref{fig:fig10} shows how local density inversions in the NESS
develop during the relaxation process if a CDW is present. For $F=V=2$, see Fig.~\ref{fig:fig10}(a),
bulk occupations evolve slowly.  In particular, $\bar n_4(t)$ first increases and subsequently decreases, while
$\bar n_5(t)$ continues to grow on the shown timescale. Their crossing eventually produces the inversion $\bar n_4<\bar n_5$ characteristic of the
stationary profile in Fig.~\ref{fig:fig5}(b).
The bulk densities are still changing near the end
of the displayed interval, illustrating that very long
relaxation times can emerge at strong tilts.
For $F=1$ and $V=3$, see Fig.~\ref{fig:fig10}(b),
several neighboring pairs develop inverted occupations.
The late-time pattern shows the ordering $\bar n_2< \bar n_3$, $\bar n_4< \bar n_5$, and
$\bar n_6< \bar n_7$, consistent with Fig.~\ref{fig:fig4}(c). These inversions reveal alternating
variations superimposed on the overall density decrease as one moves from the source to the drain electrode. 
The crossings describe how this pattern forms, but their number is not a general measure
of the extent of a CDW region. Transient crossings can also occur without stationary staggering,  see
Fig.~\ref{fig:fig9}(a).

To detect the spatial modulation of the density profile through a transport
measurement, we now retain the two end-point reservoirs
and in addition attach a mobile tip, i.e., a purely extracting reservoir weakly coupled to
a site $k$. For each possible position of the tip, the system is characterized by a different steady state, denoted by
$\rho^{\mathrm{ss},k}(\gamma_{\mathrm{tip}})$.  For simplicity, the same probe rate $\gamma_{\mathrm{tip}}$ is assumed at every position $k$. 
The outgoing tip current obeys the exact relation in Eq.~\eqref{eq:probe_density} and therefore is able to probe the local occupation of site $k$  in the presence of the tip.
To recover the density of the unperturbed ($\gamma_{\rm tip}=0$) end-point
setup, the probe should cause only minimal changes in
that state. Let $\bar n^{(0)}_j=\mathrm{Tr}(n_j \rho^{\mathrm{ss}})$
denote the onsite occupations for $\gamma_{\rm tip}=0$. If the
unperturbed NESS is unique, we have the relations
\begin{equation}
\frac{J_k}{\gamma_{\mathrm{tip}}} =\bar n_k^{(0)}+O(\gamma_{\mathrm{tip}}),
\quad
\lim_{\gamma_{\mathrm{tip}}\to0}\frac{J_k}{\gamma_{\mathrm{tip}}} =\bar n_k^{(0)}.
\label{eq:cdw_weak_probe}
\end{equation}
These relations are valid both for the time-dependent currents and in the steady state limit.
A spatial probe scan of the extraction current can therefore reconstruct the unperturbed density
profile in the weak-probe limit. Having small ratios $\gamma_{\rm tip}/\Gamma_1$ or $\gamma_{\rm tip}/\gamma_L$ between
the probe rate and the end-point rates is useful 
but does not by itself guarantee negligible disturbance by the probe tip. In particular,
if bulk relaxation is very slow and/or the end-point rates are so strong that the system is driven into a quantum Zeno regime, caution has to be exercised \cite{Misra_1977,Benenti_2009,Nava_2021}.

\begin{figure}
\centering
\includegraphics[width=0.75\linewidth]{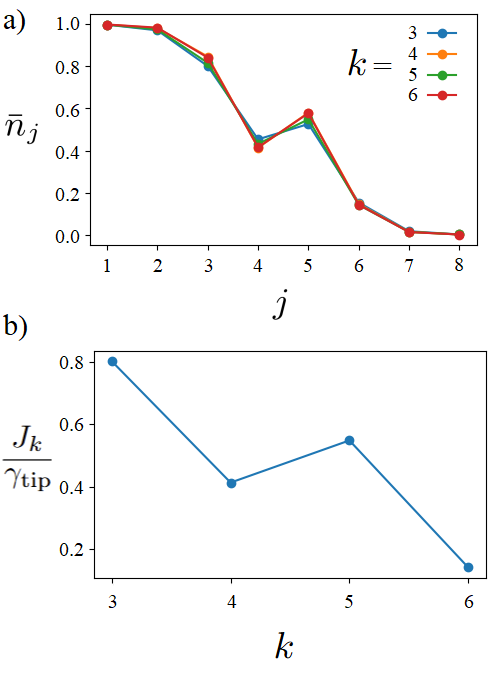}
\caption{Scanning an interacting NESS with a third extraction reservoir (mobile tip) for $L=8$, $\Gamma_1=\gamma_L=J=1$, $F=V=2$, with the mobile tip attached separately at site
$k=3,4,5,6$ with rate $\gamma_{\mathrm{tip}}=0.001$.
(a) Stationary site occupancies for the four tip positions.
(b) Outgoing stationary tip current as a function of  the tip position $k$, illustrating its proportionality to the density profile of the unperturbed ($\gamma_{\rm tip}=0$) chain.}
\label{fig:fig11}
\end{figure}

Figure~\ref{fig:fig11}(a) shows the steady-state occupancies for different positions of the tip. For weak coupling, similar profiles for the four probe positions are recovered, with the central inversion remaining visible. Their shape also resembles the unperturbed end-point profile in Fig.~\ref{fig:fig5}(b).
showing that the modulation survives
the added extraction channel.
The probe-current pattern in Fig.~\ref{fig:fig11}(b)
likewise contains a central increase from $k=4$ to
$k=5$, consistent with the density inversion.
According to Eq.~\eqref{eq:cdw_weak_probe}, the
unperturbed contrast can be obtained from separately
prepared scans as
\begin{equation}
\bar n_5^{(0)}-\bar n_4^{(0)}
=\lim_{\gamma_{\mathrm{tip}}\to0}
\frac{J_5-J_4}{\gamma_{\mathrm{tip}}}.
\label{eq:cdw_probe_contrast}
\end{equation}
The weak mobile reservoir thus provides a local density
readout through particle extraction and complements
the end-point transport measurements. The latter
characterize transmission across the chain, whereas
the scanning probe resolves the spatial modulation
within the driven state.

\section{Conclusions}
\label{sec5}

In this paper, we have investigated how local particle reservoirs
can reveal the spatial structure of transport and
density profiles in tilted interacting fermionic
chains. Within a Lindblad description, we have proposed and studied
two complementary measurement schemes: a penetration length measurement,
obtained from transport through segments of different
lengths, and a local density readout, obtained by
scanning a weak extraction reservoir across an
already driven chain.

For chains coupled to injecting and extracting
end-point reservoirs, the stationary current
defines the distance over which transmission
remains detectable at a specified threshold.
The resulting penetration length, obtained by varying the total
chain length while keeping the electric field, the interaction strength, the hopping amplitude,
and the contact rates fixed, decreases with increasing tilt and closely
follows the equilibrium density-crossover scale  introduced in
Ref.~\cite{Boidi_2025}
over the parameter range studied. This correspondence
reflects the particle confinement underlying both observables.
Interactions can extend the detectable transmission
range by compensating field-induced energy costs,
but can also suppress transport through the
formation of strongly polarized configurations.

The comparison between the NESS and a thermal TESS
benchmark shows that similar density profiles can
arise from different state-selection mechanisms.
At strong tilt, the two profiles can nearly
coincide, whereas at weaker tilt, the boundary
drive can favor polarization over the alternating CDW-type
occupations selected in equilibrium. A symmetry
analysis can relate driven profiles resembling
ground-state densities to profiles characteristic
of the upper spectral edge under simultaneous
reversal of the field and interaction; for details, see Appendix \ref{symmetries}.
These correspondences concern local occupations
and do not imply thermalization of the full
NESS.

A third, mobile extraction reservoir provides
direct access to spatial density modulations.
Its outgoing current equals the extraction rate
times the local occupation in the presence of
the probe. This relation is exact and also holds in the interacting case.
In the weak-coupling limit, the normalized current
recovers the unperturbed NESS density, allowing
neighboring density contrasts to be measured
through particle extraction. The finite-chain
results illustrate how this readout can resolve
staggered profiles associated with CDW precursors.
The relaxation dynamics also show that bulk
occupations can continue to evolve after the
boundary currents have nearly stabilized, making
the preparation time an essential part of the
density-measurement protocol.

These results establish local reservoirs 
both as a means of driving transport and as a probe
of spatial structures in open tilted quantum chains.
Implementations with spatially controlled
particle injection and loss could thereby
access both the range of transmission and
the density modulations of an interacting NESS.

\begin{acknowledgments}  
R.E. and A.N. acknowledge funding by the Deutsche Forschungsgemeinschaft (DFG, German Research Foundation) under Projektnummer 277101999 -- TRR 183 (project B02) and under Germany's Excellence Strategy -- Cluster of Excellence
Matter and Light for Quantum Computing (ML4Q) EXC 2004/2 – 390534769. 
C.R.P.~acknowledges financial support by CONICET of Argentina through Grant
No.~PIP 2022-75842.
\end{acknowledgments}

\section*{Data availability}

The data underlying the figures in this work will be made available at Zenodo.

\appendix
\section{Reservoir coupling in the energy eigenbasis}
\label{appA}

In this Appendix, we derive the weak-contact expressions
for the stationary orbital occupations and particle
current in the noninteracting chain, and clarify their
range of validity. We then discuss how interactions
modify this picture through transitions between
many-body states and local compensation of the
field-induced energy cost.

For $V=0$, diagonalization of the single-particle Hamiltonian gives
\begin{equation}
H=\sum_\alpha\epsilon_\alpha d_\alpha^\dagger d_\alpha,
\qquad c_j=\sum_\alpha\phi_\alpha(j)d_\alpha.
\label{eq:app_orbital_basis}
\end{equation}
The contribution of a reservoir at site $j$ is
\begin{equation}
\mathcal L_j\rho=g_j f_j\mathcal D[c_{j}^\dagger]\rho
+g_j(1-f_j)\mathcal D[c_{j}]\rho.
\label{eq:app_local_reservoir}
\end{equation}
Changing the basis does not by itself make the orbitals independent. For example, the loss term becomes
\begin{equation}
\mathcal D[c_{j}]\rho
=\sum_{\alpha,\beta}\phi_\alpha(j)\phi_\beta^*(j)
\left(d_\alpha \rho d_\beta^\dagger
-\frac12 \{d_\beta^\dagger d_\alpha,\rho\}\right),
\label{eq:app_orbital_dissipator}
\end{equation}
which contains terms with $\alpha\ne\beta$ even though $H$ is diagonal. In the interaction picture, these terms oscillate at
$\epsilon_\alpha-\epsilon_\beta$. For sufficiently weak contact rates, $g_j \to 0$, at fixed nonzero single-particle level spacings, we can discard these terms when the relevant energy differences are resolved on the dissipative time scale \cite{DAbbruzzo_2021}.  Degenerate or insufficiently resolved levels should instead be treated together, retaining their coherences.
Under this secular approximation,
\begin{equation}
\mathcal L_j^{\mathrm{sec}}\rho
=\sum_\alpha g_j |\phi_\alpha(j)|^2
\left[f_j\mathcal D[d_\alpha^\dagger]\rho
+(1-f_j)\mathcal D[d_\alpha]\rho\right].
\label{eq:app_secular_generator}
\end{equation}
The contact amplitudes therefore enter as squared overlaps. With
$\nu_\alpha=\langle d_\alpha^\dagger d_\alpha\rangle$, injection into orbital $\alpha$ contributes
$g_j f_j|\phi_\alpha(j)|^2(1-\nu_\alpha)$, whereas extraction
contributes $-g_j(1-f_j)|\phi_\alpha(j)|^2\nu_\alpha$.
For the end-point setup, defining $a_\alpha$ and $b_\alpha$ in Eq.~\eqref{eq:modal_occupation}, 
we obtain the closed population equation
\begin{equation}
\dot\nu_\alpha=a_\alpha f_1+b_\alpha f_L
-(a_\alpha+b_\alpha)\nu_\alpha.
\label{eq:app_population_rate}
\end{equation}
For $a_\alpha+b_\alpha>0$, its steady-state solution is given by the occupations $\nu_\alpha^{\rm ss}$ in Eq.~\eqref{eq:modal_occupation}
while the time evolution reads
\begin{align}
\nu_\alpha(t)-\nu_\alpha^{\mathrm{ss}}
&=[\nu_\alpha(0)-\nu_\alpha^{\mathrm{ss}}]
e^{-(a_\alpha+b_\alpha)t}.
\label{eq:app_modal_relaxation}
\end{align}
We note that the stationary occupations depend on the ratio of the contact
couplings, while the relaxation time depends on their sum.
An orbital can therefore have a well-defined stationary occupation and still require a very long relaxation time.
The net current injected by the left reservoir into orbital $\alpha$ and that extracted by the right reservoir are, respectively, given by
\begin{equation}
J_{1,\alpha}=a_\alpha(f_1-\nu_\alpha), \quad
J_{\alpha,L}=b_\alpha(\nu_\alpha-f_L).
\end{equation}
These currents are equal in the stationary state. 

Next, if an orbital satisfies $\phi_\alpha(j)=0$ at every reservoir
site, its number operator commutes with $H$ and with all local
jump operators. Hence
\begin{equation}
\frac{d}{dt}\langle d_\alpha^\dagger d_\alpha\rangle=0
\label{eq:app_dark_conservation}
\end{equation}
holds under the full local Lindblad equation; no secular approximation is required for this result. Such an orbital cannot be
populated (or emptied) from the vacuum by repeated injections (or extractions). 
For the finite open nearest-neighbor chain, however, an orbital cannot satisfy this condition at an end-point. Its single-particle eigenvalue equation reads
\begin{equation}
-J\phi_\alpha(j-1)+\mu_j\phi_\alpha(j)
-J\phi_\alpha(j+1)=\epsilon_\alpha\phi_\alpha(j),
\end{equation}
with the absent terms omitted at the boundaries. If
$\phi_\alpha(1)=0$, the equation at site 1 gives
$J\phi_\alpha(2)=0$. For $J\ne0$, iteration implies that every
component vanishes, contradicting normalization. The same argument applies at site $L$ and to unequal but nonzero bond hoppings. Consequently, strong localization creates weakly coupled orbitals in this geometry, rather than exact dark orbitals. Long-time initial-state memory over a finite observation time can arise from these small couplings without implying multiple stationary states. Note that, within the bulk of the chain, nodal points with zero overlaps on a given site are allowed for some eigenstates. A mobile tip can thus be blind to some eigenstates.

We now turn to the interacting case and describe the local energy competition mechanism.
For $V\ne0$, the reservoirs connect many-body eigenstates
rather than independent single-particle orbitals. If
$|a,N\rangle$ denotes an eigenstate with $N$ particles,
injection at site $j$ can populate an $(N+1)$-particle
eigenstate $|b,N+1\rangle$ through the matrix element
\begin{equation}
M_{ba}^{(j)}=\langle b,N+1|c_{j}^\dagger|a,N\rangle.
\label{eq:app_manybody_matrix_element}
\end{equation}
If this matrix element vanishes, the direct transition is forbidden. The state may nevertheless be reached
through other sequences of allowed transitions. In a weak-contact regime, where coherences can be neglected
in the population dynamics, the injection rate is given by
\begin{equation}
W_{b,N+1\leftarrow a,N}^{(j)}
=g_{j}f_{j}|M_{ba}^{(j)}|^2,
\end{equation}
with an analogous expression for extraction.
The stationary populations then follow from coupled
balance equations, rather than from independent
orbital occupations.

To understand how interactions can facilitate transport,
consider a particle hopping to the right on an interior
bond $(j,j+1)$. Initially, $n_j=1$ and $n_{j+1}=0$.
The hop increases the potential energy by $F$, but also
changes the interaction with the neighboring particles.
For the interaction
$V\sum_j(n_j-\tfrac12)(n_{j+1}-\tfrac12)$,
the total change in diagonal energy is
\begin{equation}
\Delta E=F+V(n_{j+2}-n_{j-1}),
\qquad 2\le j\le L-2.
\label{eq:app_hop_detuning}
\end{equation}
The interaction on the bond $(j,j+1)$ is unchanged since the occupied and empty sites simply exchange positions, while
the two neighboring bonds give the additional contribution $V(n_{j+2}-n_{j-1})$.
For example, the local rearrangement
$|1100\rangle\longleftrightarrow|1010\rangle$
moves a particle uphill while reducing its interaction
energy. The two configurations, with identical occupations
elsewhere, differ in diagonal energy by $F-V$.
Their mixing is therefore favored near $F=V$.
Keeping only these two configurations gives, after
subtracting a common energy,
\begin{equation}
H_{\mathrm{pair}}=
\begin{pmatrix}
0&-J\\
-J&F-V
\end{pmatrix}.
\label{eq:app_pair_hamiltonian}
\end{equation}
The hopping replaces the crossing at $F=V$ by an
avoided crossing. This illustrates how interactions
can help a particle overcome the potential gradient.
In the full chain, other configurations and the reservoirs
determine whether such mixing produces enhanced transport.
This interpretation is related to the conduction
resonances discussed in Ref.~\cite{Mendoza_2024}.
The local condition $F\simeq V$  illustrates one
possible energy compensation mechanism. Naturally, it does not determine
all resonances of the many-body system.

The role of coherences follows directly by expressing
the stationary current in the energy eigenbasis,
\begin{equation}
J^{\mathrm{ss}}=\sum_{a,b}
\langle a|\rho^{\mathrm{ss}}|b\rangle
\langle b| J_{j,j+1}|a\rangle,
\label{eq:app_manybody_current}
\end{equation}
where the labels include all particle-number sectors. Enhanced transport requires both a coherence and a nonzero current matrix element between the states involved. Near closely spaced
levels, these coherences must be retained, so population
rates alone need not describe the stationary transport.

A complementary point concerns the strength of particle
redistribution at large $F$. In the Wannier-Stark basis,
some interaction terms only shift the energies of
occupation configurations. For example,
$U_{\alpha,\beta}\hat\nu_\alpha\hat\nu_\beta$, with
$\hat\nu_\alpha=d_\alpha^\dagger d_\alpha$, changes the
energy when both orbitals are occupied but cannot
change either occupation. Other interaction terms
transfer particles between orbitals. Indeed a density-assisted hopping term
\begin{equation}
T_{\alpha\beta;\gamma}
d_\alpha^\dagger d_\beta\hat\nu_\gamma+\mathrm{H.c.}
\end{equation}
can transfer a particle from orbital $\beta$ to orbital
$\alpha$ when orbital $\gamma$ is occupied. Pair-transfer
terms of the form
\begin{equation}
P_{\alpha\beta;\gamma\delta}
d_\alpha^\dagger d_\beta^\dagger d_\delta d_\gamma
+\mathrm{H.c.}
\end{equation}
instead transfer two particles between orbitals.
All these terms arise from expressing the original
density-density interaction in the Wannier-Stark basis.
Their coefficients depend on products of orbital
amplitudes at neighboring lattice sites.
As shown in Ref.~\cite{Krajewski_2025}, for
nearest-neighbor density-density interactions, these
transfer contributions become small as the field
increases at fixed $V$.
This is intuitive because strongly localized
Wannier-Stark orbitals approach individual lattice
sites, where the interaction is diagonal in the
occupations. Interactions can therefore substantially
shift configuration energies while inducing only
slow redistribution between orbitals.

These observations distinguish two ingredients of
interaction-assisted transport: a favorable energy
difference and a coupling that connects the relevant
configurations. The reservoirs determine how those
configurations are populated.

\section{Symmetries}
\label{symmetries}

Here, we discuss transformations of the
Hamiltonian in Eq.~\eqref{eq:Ham} and their consequences for its spectrum,
the eigenstate density profiles, and the stationary
solutions of the Lindblad equation with end-points reservoirs. We distinguish transformations relating different parameters from
symmetries at fixed parameters.
  
Let us consider the particle-hole transformation
$\mathcal P$, the staggered gauge transformation
$\mathcal G$, and spatial inversion about the center
$\mathcal I$. These unitary operators act as
\begin{align}
\mathcal P c_j\mathcal P^{-1}
&=(-1)^j c_j^\dagger,
&
\mathcal P c_j^\dagger\mathcal P^{-1}
&=(-1)^j c_j,
\nonumber\\
\mathcal G c_j\mathcal G^{-1}
&=(-1)^j c_j,
&
\mathcal G c_j^\dagger\mathcal G^{-1}
&=(-1)^j c_j^\dagger,
\label{eq:sym_operators}\\
\mathcal I c_j\mathcal I^{-1}
&=c_{L+1-j},
&
\mathcal I c_j^\dagger\mathcal I^{-1}
&=c_{L+1-j}^\dagger.
\nonumber
\end{align}
Consequently,
\begin{align}
\mathcal P n_j\mathcal P^{-1}
&=1-n_j,
&
\mathcal P c_j^\dagger c_{j+1}\mathcal P^{-1}
&=c_{j+1}^\dagger c_j,
\nonumber\\
\mathcal G n_j\mathcal G^{-1}
&=n_j,
&
\mathcal G c_j^\dagger c_{j+1}\mathcal G^{-1}
&=-c_j^\dagger c_{j+1},
\label{eq:sym_bilinears}\\
\mathcal I n_j\mathcal I^{-1}
&=n_{L+1-j},
&
\mathcal I c_j^\dagger c_{j+1}\mathcal I^{-1}
&=c_{L+1-j}^\dagger c_{L-j}.
\nonumber
\end{align}

Under the particle-hole transformation,
\begin{equation}
\mathcal P H(J,F,V)\mathcal P^{-1}
=H(J,-F,V).
\label{eq:sym_PH_H}
\end{equation}
Thus, starting from
\begin{equation}
H(J,F,V)|\psi_E\rangle=E|\psi_E\rangle,
\end{equation}
we obtain
\begin{align}
\mathcal P H(J,F,V)\mathcal P^{-1}
\left(\mathcal P|\psi_E\rangle\right)
&=E\left(\mathcal P|\psi_E\rangle\right),
\nonumber\\
H(J,-F,V)\left(\mathcal P|\psi_E\rangle\right)
&=E\left(\mathcal P|\psi_E\rangle\right).
\label{eq:sym_PH_eigenstate}
\end{align}
Reversing $F$ therefore preserves the full energy
spectrum, while the mapped eigenstate exchanges
particles and holes
\begin{align}\label{eq:sym_PH_density}
    \langle \psi_E | n_j | \psi_E \rangle & = C_{j,j} \nonumber \\
    \langle \psi_E | \mathcal P^{-1} n_j \mathcal P | \psi_E \rangle & = 1- C_{j,j}.
\end{align}
In particular, $\mathcal P$ maps the ground-state
manifold of $H(J,F,V)$ onto that of $H(J,-F,V)$.
Since
\begin{equation}
\mathcal P N\mathcal P^{-1}=L-N,
\qquad N=\sum_j n_j,
\end{equation}
this transformation relates the $N$-particle and
$(L-N)$-particle sectors. It acts within a fixed
particle-number sector only at half filling.

Under the staggered gauge transformation,
\begin{equation}
\mathcal G H(J,F,V)\mathcal G^{-1}
=H(-J,F,V)
=-H(J,-F,-V).
\label{eq:sym_G_H}
\end{equation}
Accordingly,
\begin{align}
\mathcal G H(J,F,V)\mathcal G^{-1}
\left(\mathcal G|\psi_E\rangle\right)
&=E\left(\mathcal G|\psi_E\rangle\right),
\nonumber\\
H(J,-F,-V)\left(\mathcal G|\psi_E\rangle\right)
&=-E\left(\mathcal G|\psi_E\rangle\right).
\label{eq:sym_G_eigenstate}
\end{align}
Simultaneously reversing $F$ and $V$ therefore
reverses the spectrum. The mapped eigenstates have
opposite energies but identical site occupancies
\begin{align}\label{eq:sym_G_density}
    \langle \psi_E | n_j | \psi_E \rangle & = C_{j,j} \nonumber \\
    \langle \psi_E | \mathcal G^{-1} n_j \mathcal G | \psi_E \rangle & = C_{j,j}.
\end{align}
In particular, the ground state of $H(J,-F,-V)$
is obtained by applying $\mathcal G$ to a
highest-energy state of $H(J,F,V)$.

Finally, spatial inversion gives
\begin{equation}
\mathcal I H(J,F,V)\mathcal I^{-1}
=H(J,-F,V),
\label{eq:sym_I_H}
\end{equation}
and hence
\begin{equation}
H(J,-F,V)\left(\mathcal I|\psi_E\rangle\right)
=E\left(\mathcal I|\psi_E\rangle\right).
\end{equation}
Its action on the density profile is a reflection
\begin{equation}
\langle\psi_E|\mathcal I^{-1}n_j\mathcal I
|\psi_E\rangle
=C_{L+1-j,L+1-j}.
\label{eq:sym_I_density}
\end{equation}

Combining spatial inversion and particle-hole
conjugation gives a symmetry at fixed parameters,
\begin{equation}
(\mathcal I\mathcal P)H(J,F,V)
(\mathcal I\mathcal P)^{-1}
=H(J,F,V).
\label{eq:sym_IP_H}
\end{equation}
Therefore,
\begin{align}
H(J,F,V)|\psi_E\rangle
&=E|\psi_E\rangle,
\nonumber\\
H(J,F,V)\left(\mathcal I\mathcal P|\psi_E\rangle\right)
&=E\left(\mathcal I\mathcal P|\psi_E\rangle\right).
\label{eq:sym_IP_eigenstate}
\end{align}
The density of the transformed state satisfies
\begin{equation}
\langle\psi_E|(\mathcal I\mathcal P)^{-1}
n_j(\mathcal I\mathcal P)|\psi_E\rangle
=1-C_{L+1-j,L+1-j}.
\label{eq:sym_IP_density}
\end{equation}
This symmetry does not generally require a
degeneracy. The transformed state may coincide
with the original state up to a phase. For a
nondegenerate eigenvalue,
\begin{equation}
\mathcal I\mathcal P|\psi_E\rangle
=e^{i\theta_E}|\psi_E\rangle,
\end{equation}
which implies
\begin{equation}
C_{j,j}+C_{L+1-j,L+1-j}=1.
\label{eq:sym_complementary_density}
\end{equation}
Thus the density deviations from half filling
are antisymmetric about the chain center,
\begin{equation}
C_{j,j}-\frac12
=-\left(C_{L+1-j,L+1-j}-\frac12\right).
\end{equation}
If the transformed state is linearly independent
of the original one, the two states are degenerate.
This necessarily occurs for a state with a definite
particle number $N\neq L/2$, because its partner
belongs to the distinct sector $L-N$. At half
filling, however, no degeneracy follows from
$\mathcal I\mathcal P$ alone. Within a degenerate
energy subspace, arbitrary linear combinations
need not individually obey
Eq.~\eqref{eq:sym_complementary_density}.

Other combinations give useful spectral relations.
For example,
\begin{align}
(\mathcal P\mathcal G)H(J,F,V)
(\mathcal P\mathcal G)^{-1}
&=-H(J,F,-V),
\nonumber\\
(\mathcal I\mathcal G)H(J,F,V)
(\mathcal I\mathcal G)^{-1}
&=-H(J,F,-V).
\label{eq:sym_combined_spectral}
\end{align}
At $V=0$, both relations become anticommutation
relations,
\begin{equation}
\{\mathcal P\mathcal G,H(J,F,0)\}
=\{\mathcal I\mathcal G,H(J,F,0)\}=0.
\end{equation}
For either transformation $U$, an eigenstate
$|\psi_E\rangle$ with energy $E$ is mapped to
an eigenstate with energy $-E$:
\begin{equation}
H(J,F,0)\bigl(U|\psi_E\rangle\bigr)
=-E\bigl(U|\psi_E\rangle\bigr).
\end{equation}
The spectrum is therefore symmetric about zero.
The transformation $\mathcal I\mathcal G$
preserves particle number and establishes this
pairing within each particle-number sector,
whereas $\mathcal P\mathcal G$ maps the
$N$-particle sector onto the $(L-N)$-particle
sector.

The above relations have consequences for the thermal density profiles.
At chemical potential $\mu=0$, the thermal state is
\begin{equation}
\rho_{\mathrm{th}}(F,V;\beta)
=\frac{e^{-\beta H(J,F,V)}}{Z(F,V;\beta)}.
\end{equation}
Equations~\eqref{eq:sym_PH_H} and
\eqref{eq:sym_I_H} imply
\begin{align}
\rho_{\mathrm{th}}(-F,V;\beta)
&=\mathcal P\rho_{\mathrm{th}}(F,V;\beta)
\mathcal P^{-1},
\nonumber\\
&=\mathcal I\rho_{\mathrm{th}}(F,V;\beta)
\mathcal I^{-1}.
\end{align}
Consequently,
\begin{align}
\bar n_j^{\mathrm{th}}(-F,V)
&=1-\bar n_j^{\mathrm{th}}(F,V),
\nonumber\\
&=\bar n_{L+1-j}^{\mathrm{th}}(F,V).
\label{eq:sym_thermal_F}
\end{align}
These relations remain valid in the $T\to0$ limit 
if the zero-temperature thermal state is the equal statistical mixture
of the degenerate ground-state manifold.

The simultaneous reversal of $F$ and $V$ has a
different consequence,
\begin{equation}
\rho_{\mathrm{th}}(-F,-V;\beta)
=\mathcal G\rho_{\mathrm{th}}(F,V;-\beta)
\mathcal G^{-1}.
\label{eq:sym_thermal_negative_beta}
\end{equation}
The negative inverse temperature on the right-hand
side is a way of expressing the reversed ordering
of the finite spectrum. As $\beta\to+\infty$, it
selects the highest-energy manifold of the original
Hamiltonian. Therefore,
\begin{equation}
\bar n_j^{\mathrm{gs}}(-F,-V)
=\bar n_j^{\mathrm{max}}(F,V),
\label{eq:sym_ground_highest}
\end{equation}
where the superscripts denote ground-state ($\beta \rightarrow +\infty$) and
highest-energy-state ($\beta \rightarrow -\infty$) densities, respectively.
There is no corresponding requirement that the
two ground-state density profiles coincide.

For the open system, the transformations of the
Hamiltonian must be accompanied by  corresponding
transformations of the jump operators. With ${\cal D}$ in Eq.~\eqref{superop},
the dissipative contribution to the Lindblad equation \eqref{eq:lindbladeq} is
\begin{equation}
\mathcal L_{\mathrm D}\rho=\sum_j\left(
\Gamma_j\mathcal D[c_j^\dagger]\rho
+\gamma_j\mathcal D[c_j]\rho
\right).
\end{equation}
Since an overall phase of a jump operator leaves
its dissipator unchanged, the rate transformations
are
\begin{align}
\mathcal P:\quad
(\Gamma_j,\gamma_j)
&\longmapsto(\gamma_j,\Gamma_j),
\nonumber\\
\mathcal I:\quad
(\Gamma_j,\gamma_j)
&\longmapsto
(\Gamma_{L+1-j},\gamma_{L+1-j}),
\nonumber\\
\mathcal G:\quad
(\Gamma_j,\gamma_j)
&\longmapsto(\Gamma_j,\gamma_j).
\label{eq:sym_rate_transformations}
\end{align}
Thus $\mathcal P$ exchanges injection and extraction,
whereas $\mathcal I$ reflects the reservoir positions.
Their combination is a symmetry of the full
Lindblad generator when
\begin{equation}
\Gamma_j=\gamma_{L+1-j}
\qquad\text{for every }j.
\label{eq:sym_reservoir_condition}
\end{equation}
This condition holds for the end-point setup with
$\Gamma_1=\gamma_L$ and all other rates zero.
If the NESS is unique, it follows that
\begin{equation}
\rho^{\mathrm{ss}}
=(\mathcal I\mathcal P)\rho^{\mathrm{ss}}
(\mathcal I\mathcal P)^{-1},
\end{equation}
and therefore
\begin{equation}
\bar n_j^{\mathrm{ss}}
+\bar n_{L+1-j}^{\mathrm{ss}}=1,
\qquad
\langle N\rangle^{\mathrm{ss}}=\frac L2.
\label{eq:sym_NESS_density}
\end{equation}
The same relation holds during time evolution
only if the initial state also respects this
symmetry. In particular, an initially empty chain does not fulfill the relation since
$\mathcal I\mathcal P$ maps it onto the completely
filled chain. Moreover, an additional extraction
tip generally breaks Eq.~\eqref{eq:sym_reservoir_condition}.

There is also an exact relation between the driven
systems at $(F,V)$ and $(-F,-V)$ without changing
the reservoir rates. In the occupation-number basis,
where $H$ and the matrices $c_j$ are real,
complex conjugating the Lindblad equation gives
\begin{equation}
\frac{d\rho^*}{dt}
=+i[H(J,F,V),\rho^*]
+\mathcal L_{\mathrm D}\rho^*.
\end{equation}
We now define
\begin{equation}
\rho'(t)=\mathcal G\rho^*(t)\mathcal G^{-1}.
\label{eq:sym_GK_state}
\end{equation}
Using Eq.~\eqref{eq:sym_G_H} and the invariance of
the dissipators under $\mathcal G$, we obtain
\begin{align}
\frac{d\rho'}{dt}
&=+i[\mathcal G H(J,F,V)\mathcal G^{-1},\rho']
+\mathcal L_{\mathrm D}\rho'
\nonumber\\
&=-i[H(J,-F,-V),\rho']
+\mathcal L_{\mathrm D}\rho'.
\label{eq:sym_GK_LE}
\end{align}
Thus $\rho'(t)$ solves the Lindblad equation with
both $F$ and $V$ reversed and with the same
injection and extraction rates.
For a common initial state satisfying
\begin{equation}
\rho(0)=\mathcal G\rho(0)^*\mathcal G^{-1},
\end{equation}
as is the case for the empty chain or any state
diagonal in the occupation-number basis, this gives
\begin{equation}
\rho(t;-F,-V)
=\mathcal G\rho(t;F,V)^*\mathcal G^{-1}.
\label{eq:sym_GK_dynamics}
\end{equation}
Since the density operators are real and invariant
under $\mathcal G$,
\begin{equation}
\bar n_j(t;-F,-V)=\bar n_j(t;F,V).
\label{eq:sym_GK_density}
\end{equation}
For unique stationary states, the analogous
relation holds independently of the preparation
\begin{align}
\rho^{\mathrm{ss}}(-F,-V)
&=\mathcal G\rho^{\mathrm{ss}}(F,V)^*
\mathcal G^{-1},
\nonumber\\
\bar n_j^{\mathrm{ss}}(-F,-V)
&=\bar n_j^{\mathrm{ss}}(F,V).
\label{eq:sym_GK_NESS}
\end{align}
The stationary reservoir currents are consequently
unchanged, because they depend only on the local
occupancies and the unchanged rates. The bond
currents are unchanged as well as complex conjugation
and $\mathcal G$ each reverse the sign of the
bond-current operator, so their combined action
preserves its expectation value. In the end-point
setup,
\begin{equation}
J^{\mathrm{ss}}(-F,-V)
=J^{\mathrm{ss}}(F,V).
\label{eq:sym_GK_current}
\end{equation}
These equalities concern the densities and currents;
the density matrices are related by
Eq.~\eqref{eq:sym_GK_NESS} and need not be identical.

\begin{figure}
\centering
\includegraphics[width=0.75\linewidth]{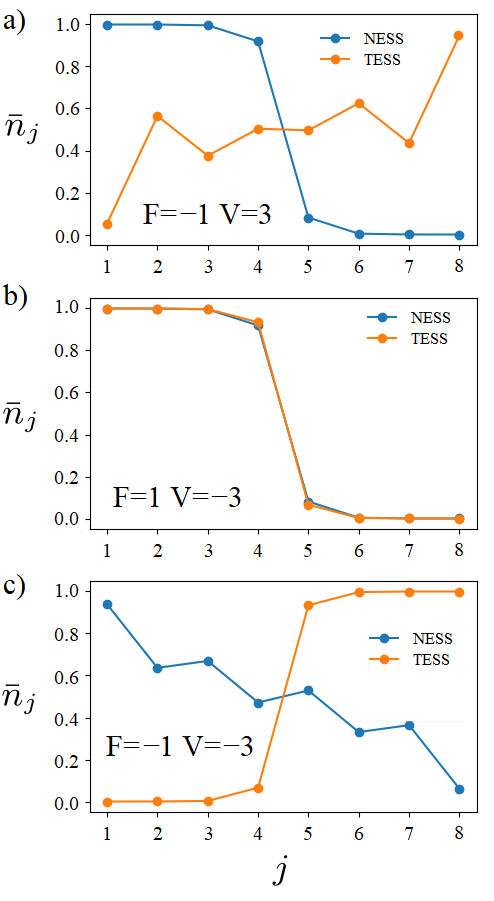}
\caption{Steady-state site occupancies for $L=8$ and $\Gamma_1=\gamma_L=J=1$
with (a) $F=-1$ and $V=3$,
(b) $F=1$ and $V=-3$,
and (c) $F=-1$ and $V=-3$.
Blue curves denote the NESS.
Orange curves denote the TESS at $\mu=0$ in
the $T\to0$ limit. }
\label{fig:fig12}
\end{figure}

\begin{figure}
\centering
\includegraphics[width=0.65\linewidth]{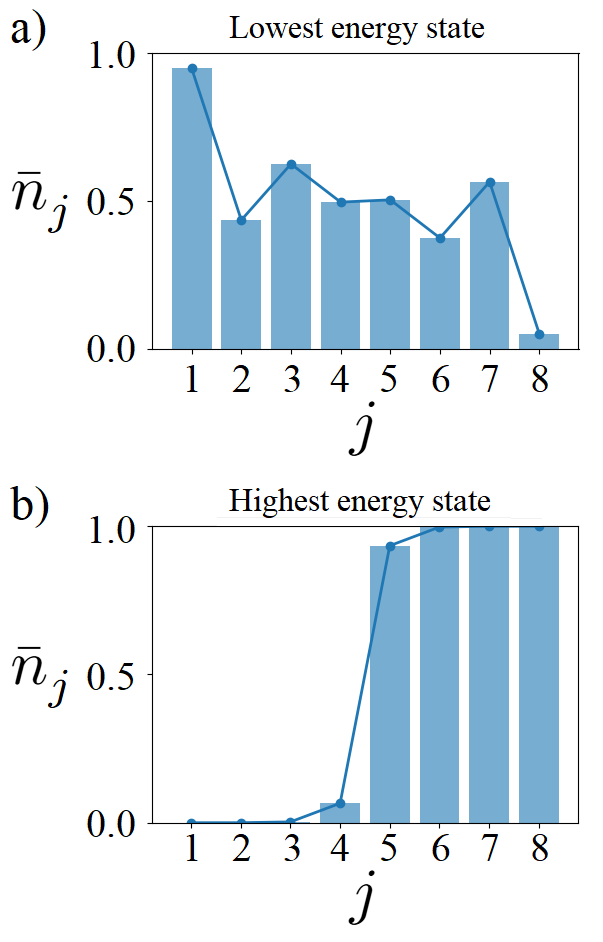}
\caption{Site occupancies of the
(a) lowest-energy and (b) highest-energy
eigenstates of the closed chain for $L=8$, $J=1$,
$F=1$, and $V=3$.
The highest-energy-state profile corresponds,
under the staggered gauge transformation, to
the ground-state profile for $F=-1$ and $V=-3$.}
\label{fig:fig13}
\end{figure}

Figure~\ref{fig:fig12} illustrates the different
consequences of these transformations on the density profiles for thermal
and driven states. In Fig.~\ref{fig:fig12}(a), reversing the
field $F$ relative to the one in Fig.~\ref{fig:fig4}(c) reflects
the thermal density profile, as required by
Eq.~\eqref{eq:sym_thermal_F}. The NESS instead
remains strongly occupied on the left side and almost
empty on the right side. Reversing the field alone
does not imply a reflection of the NESS profile if the source and
drain reservoirs are kept at their original positions.

Figures \ref{fig:fig12}(a) and (b) are related by the simultaneous
reversal of $F$ and $V$. Their NESS density
profiles therefore coincide, in agreement with
Eq.~\eqref{eq:sym_GK_NESS}, although their thermal
profiles are very different. For the attractive
interaction $(V<0)$ and positive tilt ($F>0$) in Fig.~\ref{fig:fig12}(b),
the thermal profile is strongly polarized and is close to the driven profile. 
This comparison also admits a negative-temperature interpretation. Since the staggered transformation
preserves site occupations while reversing the spectrum, see Eq.~\eqref{eq:sym_ground_highest},
the thermal profile in Fig.~\ref{fig:fig12}(b) represents the negative-temperature-limit
profile of the Hamiltonian in Fig.~\ref{fig:fig12}(a).
Together with the strong resemblance of the NESS densities in both panels, their agreement
in Fig.~\ref{fig:fig12}(b) implies
\begin{equation}
\bar n_j^{\mathrm{NESS}}(-F,V) =
\bar n_j^{\mathrm{NESS}}(F,-V)
\simeq
\bar n_j^{\mathrm{max}}(-F,V).
\end{equation}
The boundary drive therefore stabilizes a density
profile characteristic for the upper spectral edge.
We stress that this correspondence concerns the local densities only and
does not claim that the full NESS is a
negative-temperature Gibbs state.

In Fig.~\ref{fig:fig12}(c), both signs of $F$ and $V$ are reversed relative
to $F=1$ and $V=3$ in Fig.~\ref{fig:fig4}(c).
The NESS reproduces the density profile of Fig.~\ref{fig:fig4}(c), including its weak alternating variations.
The thermal state instead has an almost empty left region and an almost filled right region.
This difference follows from the spectral reversal, its ground-state density is that of
a highest-energy state at $F=1$ and $V=3$.

Finally, Fig.~\ref{fig:fig13} makes this spectral relation explicit. The lowest-energy state in Fig.~\ref{fig:fig13}(a)
has a staggered density profile superimposed on the field-induced polarization. The highest-energy
state in Fig.~\ref{fig:fig13}(b) is instead strongly polarized
toward the right side. Its density matches the thermal
profile in Fig.~\ref{fig:fig12}(c), as predicted by
Eq.~\eqref{eq:sym_ground_highest}. Both profiles
satisfy the complementary reflection relation
in Eq.~\eqref{eq:sym_complementary_density}.

\bibliography{biblio}

\end{document}